\documentclass[a4paper,fleqn,usenatbib,useAMS]{mnras}

\usepackage{graphicx}	
\usepackage{amsmath}	
\usepackage{amssymb}	
\usepackage{multicol}        
\usepackage{bm}		
\usepackage{pdflscape}	

\usepackage[T1]{fontenc}
\usepackage{ae,aecompl}

\usepackage{newtxtext,newtxmath}

\title[ExoEarth's stratosphere circulation]{Stratosphere circulation on tidally locked ExoEarths}
\author[L. Carone, R. Keppens and L. Decin]{L. Carone$^{1}$\thanks{E-mail:
carone@mpia.de (LC)},  R.
Keppens$^{2}$, L. Decin$^{3}$ and Th. Henning$^{1}$\\
$^{1}$ Max-Planck-Institute for Astronomy, Koenigsstuhl 17, 69117 Heidelberg, Germany\\
$^{2}$Centre for mathematical Plasma Astrophysics, Department of Mathematics, KU Leuven, Celestijnenlaan 200B, 3001 Leuven, Belgium\\
$^{3}$Instituut voor Sterrenkunde, KU Leuven, Celestijnenlaan 200D, 3001 Leuven, Belgium}

\date{TBD}

\pubyear{2017}

\begin{document}
\label{firstpage}
\pagerange{\pageref{firstpage}--\pageref{lastpage}} \pubyear{TBD}

\maketitle

\begin{abstract}
Stratosphere circulation is important to interpret abundances of photo-chemically produced compounds like ozone that we aim to observe to assess habitability of exoplanets. We thus investigate a tidally locked ExoEarth scenario for TRAPPIST-1b, TRAPPIST-1d, Proxima Centauri~b and GJ 667 C~f with a simplified 3D atmosphere model and for different stratospheric wind breaking assumptions. 

These planets are representatives for different circulation regimes for orbital periods: $P_{orb}=1-100$~days.
The circulation of exoplanets with $P_{orb} \leq $ 25~days can be dominated by the standing tropical Rossby wave in the troposphere and also in the stratosphere:  It leads to a strong equatorial eastward wind jet and to 'Anti-Brewer-Dobson'-circulation that confines air masses to the stratospheric equatorial region. Thus, the distribution of photo-chemically produced species and aerosols may be limited to an 'equatorial transport belt'. In contrast, planets with $P_{orb}>25$~days, like GJ~667~C~f, exhibit efficient thermally driven circulation in the stratosphere that allows for a day side-wide distribution of air masses.

The influence of the standing tropical Rossby waves on tidally locked ExoEarths with $P_{orb} \leq 25$~days can, however, be circumvented with deep stratospheric wind breaking alone - allowing for equator-to-pole transport like on Earth. For planets with $3 \leq P_{orb} \leq 6$~days, the extratropical Rossby wave acts as an additional safe-guard against the tropical Rossby wave in case of shallow wind breaking. Therefore, TRAPPIST-1d is less prone to have an equatorial transport belt in the stratosphere than Proxima~Centauri~b. 

Even our Earth model shows  an equatorial wind jet, if stratosphere wind breaking is inefficient.
\end{abstract}



\section{Introduction}
\label{Introduction}
\begin{figure*}
\includegraphics[width=0.85\textwidth]{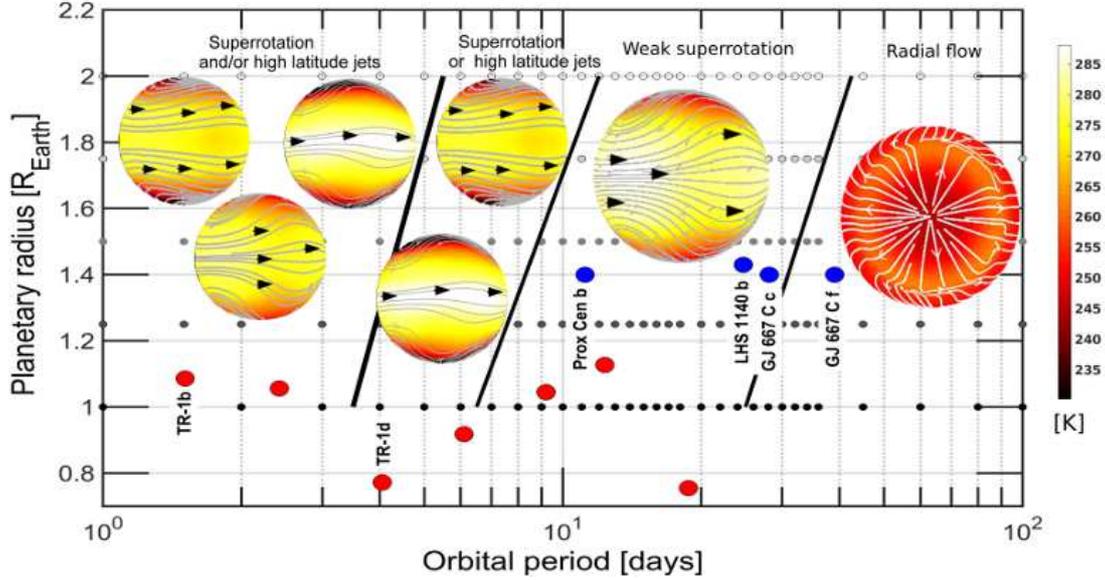}
\caption{Different troposphere circulation states identified by \citet{Carone2015} for tidally locked terrestrial planets with respect to orbital period and planetary size, assuming Earth-like atmosphere and thermal forcing. Example planets show a cross section of the planet's temperature (color) and wind flow (grey lines and black arrows) at the top of the troposphere $p=225$~mbar and facing the substellar point.
Red circles denote the radius and orbital period of the TRAPPIST-1 b,c,d,e,f,g,h planets in this 'circulation state map' from left to right. Blue circles denote the  position of Proxima Centauri b, LHS~1440~b, GJ~667~C -c and -f, assuming Earth-like density for non-transiting planets. The black to white small circles denote the 3D climate simulations carried out by \citet{Carone2015}. }
\label{fig: climate}
\end{figure*}
Tidally locked planets in the habitable zone of cool red dwarf stars like the TRAPPIST-1 planets, Proxima Centauri b, LHS 1140 b and GJ 667 C f \citep{Gillon2016,Gillon2017,Anglada2013,Anglada2016,Dittmann2017,Luger2017} are our next best hope to study conditions for habitability outside of the Solar System. Proxima Centauri b is the nearest exoplanet, located only 1.295 pc away  from the Sun. The TRAPPIST-1 planets are only 12 pc away and, in addition, all seven planets transit in front of their star with relatively short orbital periods ($P_{orb}=1.5 - 18.8$~days). Thus, the TRAPPIST-1 planets are particularly suited to search for bio-signatures in diverse environments by employing, e.g., infrared spectroscopy with JWST. However, infrared spectroscopy mainly probes high atmospheric layers, which would correspond to the stratosphere and mesosphere on Earth ($p\leq 0.1$~bar or altitudes greater than 20 km, see e.g. \citet{Barstow2016, Kreidberg2016,Morley2017}). To discuss habitability, we have to reliably infer the properties of the underlying troposphere and potentially habitable surface from upper atmosphere composition derived from infrared spectra. 

The abundances of  molecular species and clouds that we can detect on rocky planets by 'far remote-sensing' are modified by the general large-scale circulation. On Venus, surface reactions produce COS and H${}_2$S that are transported upward as part of the global Hadley circulation cells. In the higher atmosphere, the sulfur-containing compounds are further processed via photo-chemistry to SO${}_2$ and sulfuric acid to form clouds. The global super-rotating flow results then in a continous cloud blanket that enshrouds Venus completely (see e.g. \citet{Prinn1987} for a review on Venus chemistry and climate). On Earth, the Brewer-Dobson-circulation in the stratosphere tends to transport the biomarker ozone from  its main production regions in the tropical upper stratosphere towards the polar lower stratosphere (e.g., \citet{Shaw2008,Brewer1949,Dobson1931}).  Also the abundances of  CH${}_4$ - another biomarker - are affected by stratospheric Brewer-Dobson-circulation \citep{Cordero2001}. On Mars, 3D circulation is instrumental for the formation of a polar ozone layer \cite{Mont2013}. In summary, 3D circulation is crucial for the understanding of the abundances and distribution of aerosols, photo-chemically produced compounds and clouds in the upper atmosphere of terrestrial, habitable exoplanets that we will characterize in the near future.

In this work, we will investigate stratosphere circulation on tidally locked habitable planets with planetary obliquity zero in the habitable zone of ultra-cool dwarf stars for a large range of relevant orbital periods ($P_{orb}=1.5 -39$~days). We further assume to first order Earth-like atmosphere composition and thermal forcing. Such planets develop highly interesting un-Earth-like circulation patterns that warrant in-depth investigations, as we already showed  for the troposphere \citep{Carone2014, Carone2015, Carone2016}. 

We  will investigate stratospheric circulation in different troposphere circulation regimes that develop for specific orbital period ranges. One of the characteristics of these circulation regimes is that  troposphere winds are more or less forced into banded or zonally confined wind structures via two different Rossby waves. The tropical Rossby wave leads to a single equatorial superrotating jet, the extratropical Rossby wave to two high latitude wind jets. The specific circulation regimes are (Figure~\ref{fig: climate}):
\begin{itemize}
\item  ultra-short orbital periods with strong mixture of equatorial superrotation and high latitude wind jets ($P_{orb}\leq 3$~days)
\item short orbital periods with dominance of either equatorial or high latitude wind jets ($P_{orb} = 3- 6$~days)
\item intermediate orbital periods with weak superrotation  ($P_{orb}=6-25$~days) and
\item long orbital periods ($P_{orb} \geq 25$~days) with no standing Rossby waves and therefore mainly radial wind flow instead of banded wind structures.
 \end{itemize}
For each circulation regime, we also investigate two different stratosphere wind breaking assumptions. We chose the following representative examples for each circulation regime: A habitable TRAPPIST-1b, TRAPPIST-1d, Proxima Centauri b and GJ 667 ~C~ f scenario. We will show that, on many tidally locked ExoEarths, planetary waves propagate vertically from the troposphere into the stratosphere with different outcomes in stratosphere circulation: either circulation is confined to an 'equatorial transport belt' or circulation favors day-side wide distribution of air mass via thermally driven equator-to-pole-wards transport. 

\section{Model}

We use the model introduced in \citet{Carone2014} for terrestrial tidally locked planets with greenhouse gas atmospheres. We extend the radiative-convective equilibrium temperatures derived there with a stratosphere extension as described below. The extended  radiative-convective equilibrium temperatures are used in the Newtonian cooling framework as thermal forcing (see Section~2.3 in  \citet{Carone2014}) in the 3D global circulation model MITgcm \citep{Adcroft2004} that solves the primitive hydro-static equations (see Section~2.1 in \citet{Carone2014}). We furthermore use a Rayleigh surface friction prescription as described in Section~2.2 \citet{Carone2014}, where we vary in the following - when appropriate - the maximum surface friction time scale $\tau_{s,fric}$ between 0.1 to 10 days to establish different circulation states for one and the same planet \citep{Carone2016}.

\subsection{Stratosphere temperature extension and vertical resolution}
We assume for the day side troposphere the equations outlined in Sect. 2.4.2 of \cite{Carone2014}. These are combined with an adaptation of the "modification of the Held and Suarez forcing to include stratospheric structure" by \cite{William1998}. More precisely, the new equilibrium temperature in the troposphere, $T_{trop}(\nu, \phi, p)$, is:
\begin{equation}
T_{trop}(\nu,\phi,p)=max \left[T_{str,min},T_{DS,s}(\nu,\phi)\left(\frac{p}{p_s}\right)^{R/c_p}\right]\label{eq: tropo},
\end{equation}
where $\nu, \phi, p$ and $p_s$ are the planetary latitude, longitude, pressure level and surface pressure, respectively. $R$ is the specific gas constant and $c_p$ the specific heat capacity of the atmosphere, respectively. $T_{str,min}$ is the minimum stratosphere temperature, where we use here the value used by \cite{William1998} for Earth, which is $T_{str,min}=200$~K. $T_{DS,s}(\nu,\phi)$ is the day side temperature as given in Equation~(23) of \citet{Carone2014}.

The extended equilibrium temperature $T_{eq}(\nu, \phi, p)$ is then defined, following \cite{William1998}:
\begin{eqnarray}
&&T_{eq}(\nu, \phi, p)= T_{trop}(\nu,\phi,p)\cdot\nonumber\\
&&\left[ min\left(1,\frac{p}{p_d}\right)^{R\gamma_d/g}+min\left(1,\frac{p}{p_i}\right)^{R\gamma_i/g}-1\right] \label{eq: strato},
\end{eqnarray}
where $R$ is the specific gas constant for the planet's atmosphere and $g$ is surface gravity. We further assume $\gamma_d=2$~K/km, $\gamma_i= -3.345$~K/km and $p_d=100$~mbar. $p_i$ defines the day side equatorial regions, where the stratosphere is heated by UV flux. $p_d$ defines to first order tropopause location in vertical pressure coordinates.

It should be noted that this description covers a situation during enhanced stellar activity, where M dwarfs can emit UV flux on the order of that emitted by the Sun \citep{Segura2005,Venot2016}. TRAPPIST-1, for example, shows indeed frequently increased activity \citep{Vida2017}.

\begin{figure}
\includegraphics[width=0.49\textwidth]{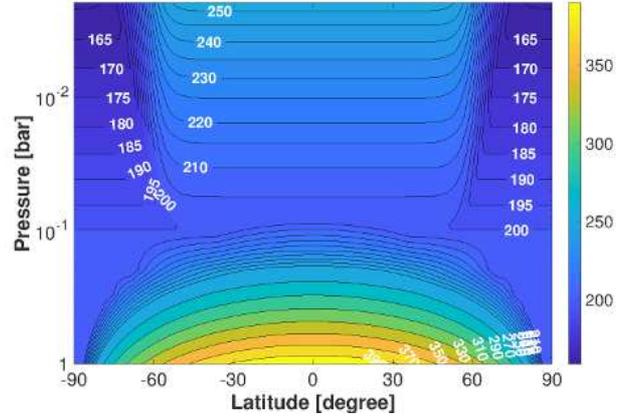}
\caption{Radiative equilibrium temperatures on the day side with stratosphere modification. Contour intervals are 5~K for temperatures colder and 20 K for temperatures hotter than 250~K.  Here, Earth radius and  gravity $g=9.81$~m/s${}^2$ is assumed.}
\label{fig: Teq}
\end{figure}

We further assume for the polar and night side regions
\begin{equation}
p_i=p_d-\left(p_d-p_{str}\right)\frac{1}{2}\left(1+tanh\left[A \left(|\zeta|-\zeta_0 \right) \right]\right),
\end{equation}
where $p_{str}=2$~mbar defines the  stratosphere model top. $A=2.65/15$ and $\zeta_0=60^{\circ}$ define the location of the polar and night side stratosphere boundary, where $\zeta$ is the stellar zenith angle, which is connected to latitude $\nu$ and longitude $\phi$ on the planet via
\begin{eqnarray}
&&cos\zeta=0 \textrm{ for } |\phi|\geq 90^{\circ} \nonumber\\
&&cos\zeta=cos\nu\cdot cos\phi \textrm{ for }|\phi|< 90^{\circ},
\end{eqnarray}
where the substellar point is located at $\phi=0,\nu=0$. We use for the remainder of this work the parameters for the troposphere temperature defined in \cite{Carone2014} for the Earth-case (see Section~\ref{sec: planet param} for more details). The equilibrium temperatures on the day side are shown in Figure~\ref{fig: Teq} for surface gravity $g=9.81$~m/s${}^2$ and Earth-like atmosphere and surface temperatures.

We retain for the horizontal resolution the original C32 cubed sphere resolution as  used in \cite{Carone2014}. This set-up yields a global resolution in longitude and latitude of $128 \times 64$ or $2.8^{\circ} \times 2.8^{\circ}$. For the vertical resolution, we use in total 30 vertical levels between 1000 mbar surface pressure and 1~mbar:  We linearly resolve the vertical extent of the atmosphere with 18 times 50 mbar levels between 1000 and 100~mbar pressure, we logarithmically resolve the atmosphere between 100 and 10~mbar with 8 levels, we further resolve linearly with 4 times 2.5 mbar levels the atmosphere between 10 to 1~mbar. This setup ensures sufficient vertical resolution for the development of the troposphere and stratosphere wind system (see also \cite{William1998}). We use a dynamical time step $\Delta t=300$~s.

\subsection{The role of wave breaking in the stratosphere}
\label{sec: strato}

In previous works (e.g. Section~2.2 in \cite{Carone2014}), we have used an upper atmosphere Rayleigh friction layer as a first order approach to mimic gravity wave breaking. This upper atmosphere friction layer was, however, mainly motivated to ensure numerical stability of the global circulation model by preventing non-physical wave reflection at the upper boundary. The upper atmosphere friction layer is prescribed as an additional forcing term $\vec{F}_v=- k_R\vec{v}$ acting on the horizontal flow, where $k_R$ is defined as
\begin{eqnarray}
k_R&=&k_{max}\left[\frac{p_{low}-p}{p_{low}}\right]^2 \quad \mathrm{if}~ p\geq p_{low}\\
k_R&=&0 \qquad\mathrm{if}~ p< p_{low}.\nonumber
\end{eqnarray}
$p_{low}$ is the lower boundary of the stratosphere wind breaking layer applied in our simulations.

In this work, upper atmosphere wind breaking is a necessary ingredient to investigate to first order the strength of wave driven circulation in the stratosphere: This type of circulation can be excited by the interaction between horizontal flow and upwards propagating waves originating from the underlying troposphere. At the same time, strong zonal flow in the stratosphere will be driven by vertically upward propagating eddies that inject momentum into the stratosphere and accelerate the mean zonal flow. Strong or weak wind breaking leads to more or less effective friction in the upper atmosphere, which regulates the acceleration of the flow;  as will be shown in this work. Wave-driven stratospheric circulation associated with strong zonal flow is indeed found on Earth in the form of the Brewer-Dobson-circulation and is associated with polar jets (see e.g. \citep{Vallis, Holton} and subsequent subsection). Such a circulation may also be found in brown dwarfs and directly imaged planets \citep{Showman2013}.

We study two cases in  this work: deep and shallow stratosphere wind breaking. In the first case $p_{low}$ is set to the lowest possible value for wind breaking: Wind breaking can take place, when planetary waves enter from the radiative-convective region into the stably stratified stratosphere region. The boundary between these atmosphere regions is located at about $p_{low}=0.1$~bar. Conversely, $p_{low}$ for the shallow wind breaking prescription was set as high up in our prescription as possible without becoming numerically unstable. These conditions were fulfilled for $p_{low}=0.005$~bar.

The value for $k_{max}$ in the deep wind breaking case is selected such that it reproduces the Earth regime. It will be shown in the next subsection, that the deep stratosphere wind breaking set-up with $p_{low}=0.1$~bar and $k_{max}=1/20$~days${}^{-1}$  is needed to reproduce qualitatively the Earth stratosphere wind and circulation regime. These values will be kept for all planets, when applying deep stratosphere wind breaking.

The value for $k_{max}$ in the shallow wind breaking case is mainly driven by the need to ensure numerical stability as $p_{low}=0.005$~bar is located very near the global circulation model top boundary; as in the original purely tropospheric set-up. Thus, $k_{max}$ is increased for the shallow wind breaking set-up and in the short orbital period regime $P_{orb}\leq 3$~days; as also discussed in \cite{Carone2015}. Interestingly, we found that for TRAPPIST-1d, a very strong $k_{max}$ is needed to ensure numerical stability. We attribute the need for an unusually high stability near the upper boundary to the complex wave behavior that can enfold in the stratosphere of this specific planet compared to all other planets, as will be shown in this study. See Table~\ref{tab: TR-param} for the specific values used for different scenarios.

\subsection{Earth benchmarking and eddy diagnostics}
\label{sec: Earth}
\begin{figure}
\includegraphics[width=0.49\textwidth]{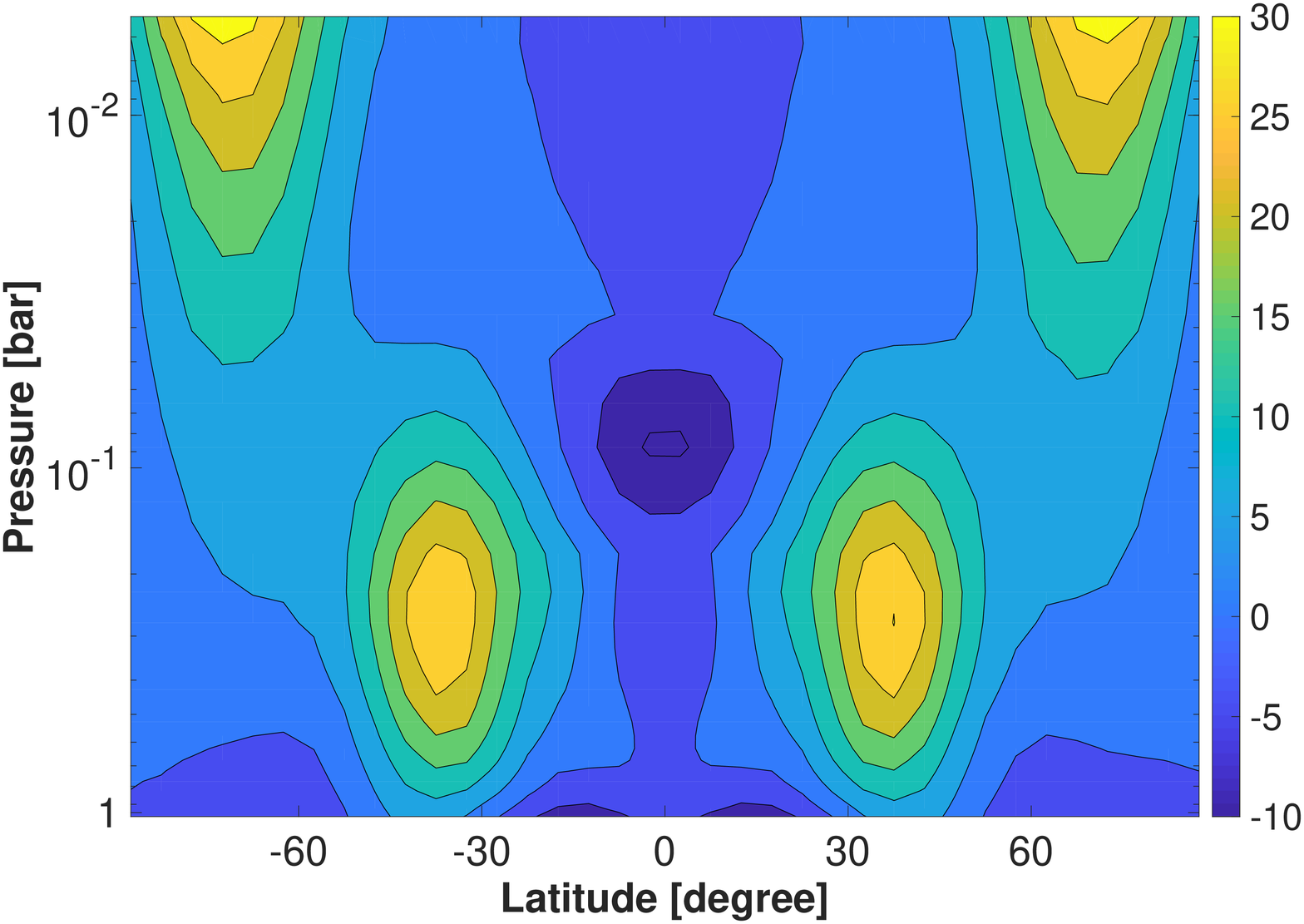}
\includegraphics[width=0.49\textwidth]{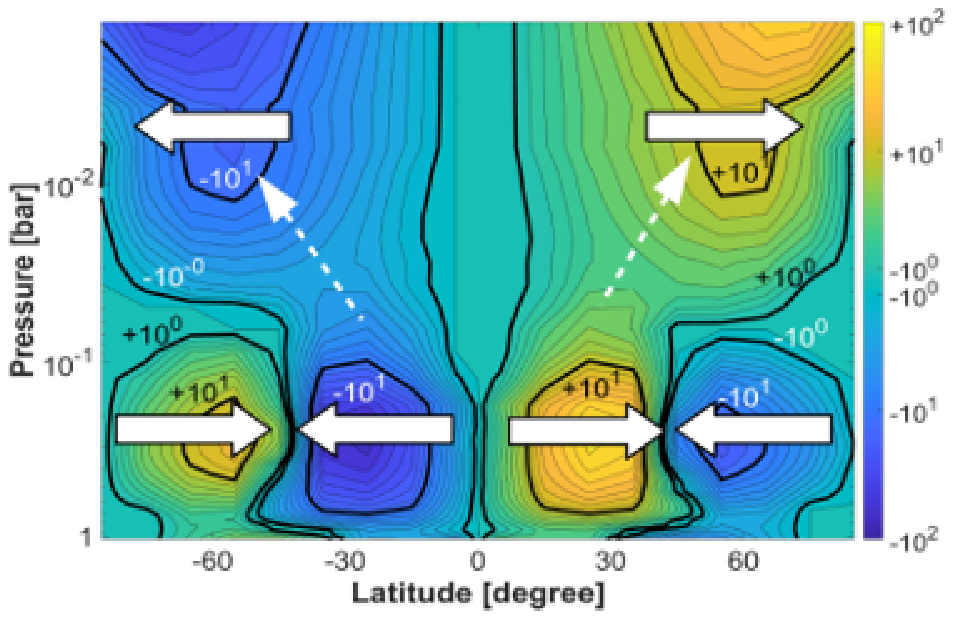}
\caption{Top: Zonally averaged zonal wind jets, averaged over 1000~days of simulation time, for the Earth benchmark setup. Contour level spacing is 5~m/s. Bottom: Zonally and time  averaged eddy momentum flux in m${}^2$/s${}^2$ for the Earth setup. Contour levels are arranged in steps of $10^{0.1}$  between $10^{2}$ and $10^0$~m${}^2$/s${}^2$ and $-10^{2}$ and $- 10^0$~m${}^2$/s${}^2$. White solid arrows denote direction of angular momentum transport. Dashed white lines denote the vertical propagation of extra tropical Rossby waves.}
\label{fig: Earth}
\end{figure}

We benchmark our model with an Earth-simulation to ensure that we reproduce qualitatively Earth-like circulation not only in the troposphere, but also in the stratosphere. The surface friction time scale is set here to the nominal value of 1~day, as defined by \cite{Held1994}, the benchmark of which we adapted for tidally locked ExoEarths in \citet{Carone2014}. We also tested slower and higher surface friction time scales ($\tau_{fric}=0.1$ and 10~days) to check if we can change the circulation state of Earth within the frame work of our model, but found no large effect in general circulation. This is in contrast to tidally locked ExoEarths were the circulation state can change drastically for different surface friction time scales \citep{Carone2016}. We find that the selected surface friction and temperature prescription used by \cite{William1998} generally reproduce the desired zonal wind and circulation structure: Mid-latitude zonal wind jets in the troposphere with wind speeds of 30 m/s,  polar wind jets in the stratosphere. A deep stratosphere wind braking description with $p_{low}=0.1$~bar and $k_{max}=1/20$~days${}^{-1}$ furthermore reproduces the observed wind speeds (Figure~\ref{fig: Earth}, upper panel). See, e.g., \cite{Manabe1968} or the review by \citet{Haynes2005} for comparison with the observed Earth zonal wind structure.

The polar zonal wind jets on Earth are strong indicators for the presence of a Brewer-Dobson-circulation in the stratosphere. Both, the polar jet as well as the Brewer Dobson circulation, are driven by extratropical Rossby waves that originate in the troposphere, propagate upwards and break in the stratosphere (see also Figure~8 in \citet{Showman2013}).

\begin{figure}
\includegraphics[width=0.49\textwidth]{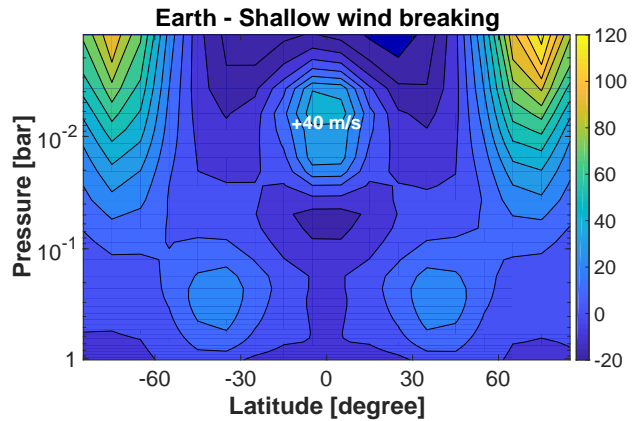}
\caption{Zonally averaged zonal wind jets, averaged over 1000~days of simulation time, for the Earth setup with shallow stratosphere wind breaking. Contour level spacing is 10~m/s. }
\label{fig: Earth_eq}
\end{figure}

Interestingly, when we test the shallow wind breaking scenario for the Earth model, that is, set $p_{low}=0.005$~bar (see Section~\ref{sec: strato}), we not only find much faster polar jets but also a 40~m/s strong equatorial wind jet in the stratosphere (Figure~\ref{fig: Earth_eq}). See also Figure~2 in \citet{Showmanbook} that illustrates the tendency for equatorial superrotation on rocky (exo-)planets. The focus of this work is tidally locked exoplanets, but we note that future work may also aim to investigate under which circumstances superrotation may appear on Earth (see also short discussion of superrotation in Earth climate models in Section~\ref{sec: summary}).

In the following, we introduce two diagnostics to identify wave activity and circulation more directly. The momentum transport by Rossby waves in our simulations can be identified by mapping the eddy momentum flux, which is defined as (see e.g. \citet{Holton}:
\begin{equation}
\left [\overline{U'V'}\right] =\left[\overline{UV}\right]-\left[\overline{U} \right]\cdot \left[\overline{V}\right].
\end{equation}
Here, $U$ is the zonal wind and $V$ is the meridional wind. Overbars denote time averaged flow, brackets denote zonally averaged flow and primes indicate the eddy part of the flow.

To identify circulation, we use the zonal average of the time averaged meridional overturning streamfunction $\overline{\Psi}$:
\begin{equation}
\overline{\Psi}=\frac{2\pi R_P}{g}\cos\nu \int_0^p \overline{V} dp',\label{eq: Psi}
\end{equation}
where $R_P$ is the radius of the planet and $g$ its surface gravity.  $\Psi$ is positive for clockwise circulation and negative for counter-clockwise circulation.

\begin{figure}
\includegraphics[width=0.49\textwidth]{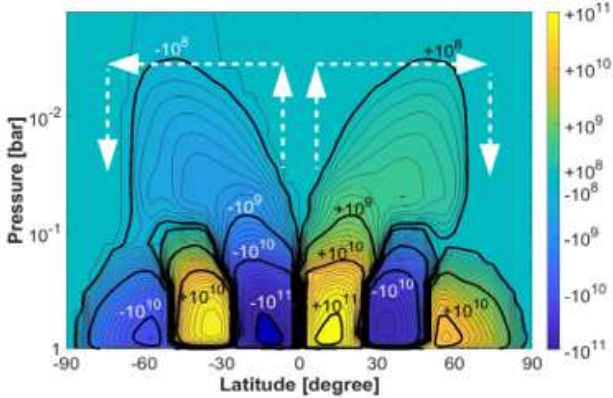}
\caption{Meridional stream function in kg/s for the Earth setup. Contour levels are arranged in steps of $10^{0.1}$ between $10^{11}$ and $10^8$ ~kg/s and $-10^{11}$ and $- 10^8$ ~kg/s, respectively.  Dashed white arrows denote direction of circulation in the stratosphere.}
\label{fig: Earth3}
\end{figure}

Figure~\ref{fig: Earth}, lower panel, shows the expected momentum transport due to Rossby waves in the troposphere and stratosphere, which explains the location of the zonal wind jets in Figure~\ref{fig: Earth}, upper panel (see also \citet{Holton}). Figure~\ref{fig: Earth3}  shows qualitatively the equator-to-pole-ward air mass transport in the stratosphere \citep{Brewer1949,Vallis}, as well as the three circulation cells per hemisphere in the troposphere (see e.g. \citet{Showmanbook,Vallis}). In the following, a combination of zonal wind, eddy momentum transport and circulation maps will be used to identify if and under which conditions thermally driven and wave driven circulation can operate in the stratosphere of tidally locked ExoEarths.


\subsection{Planetary Parameters}
\label{sec: planet param}
\begin{table*}
\caption{Planetary, atmosphere and stratosphere wind breaking parameters used in this work for simulating TRAPPIST-1b,-1d, Proxima Centauri b and GJ~667~C~f}
\begin{tabular}{c|c|c|c|c}
\hline
Planets & TRAPPIST-1b & TRAPPIST-1d  & Prox Cen b & GJ~667~C~f\\
\hline
\multicolumn{5}{|c|}{Planetary parameters} \\
\hline
Planet radius${}^a$ $R_P$ [$R_{Earth}$] & 1.086 & 0.772 & 1.4 &1.4\\
Density $\rho_P$ [$\rho_{Earth}$] & 0.62 & 1&0.71 & 1\\
Planet mass $M_P$ [$M_{Earth}$] & 0.79 & 0.46  &1.96 & 2.7\\
Surface gravity $g$ [m/s${}^2$] & 6.56  & 7.60  & 9.80&13.5\\
Orbital period [days] & 1.51  & 4.05  & 11.186 & 39.03\\
\hline
\multicolumn{5}{|c|}{atmosphere parameters} \\
\hline
Incident stellar irradiation $I_{0}$ [~W/m${}^2$] &1916${}^a$ & 1555 &900&790${}^b$ \\
albedo  & 0.5 & 0.38 & 0.3 & 0.3\\
surface pressure [bar] & 1 & 1 & 1 & 1\\
main constituent & \multicolumn{4}{|c|}{nitrogen} \\
molecular mean mass $\mu$ [g/mol] & 28 & 28 & 28 & 28 \\
specific heat capacity $c_p$ [JgK${}^{-1}$] & 1.04 & 1.04 & 1.04 & 1.04\\
average effective temperature $T_{eff}$ [K]  & 255 & 255 & 230 & 222 \\
surface optical depth ${}^c$  $\tau_s$& 0.62 & 0.62 & 0.94 & 0.94 \\
\hline
\multicolumn{5}{|c|}{stratosphere wind breaking parameters} \\
\hline
stratosphere friction timescale deep ($1/k_{max}$) [days] & 20 & 20 & 20 & 20 \\
stratosphere friction timescale shallow ($1/k_{max}$) [days] & 10 & 1 & 20 & 20 \\
\hline
\end{tabular}
\label{tab: TR-param}
\newline
${}^a$ would require a star with radius of $R_{star}=0.1 R_{Sun}$  and effective temperature $T_{star,eff}=2095$~K.\newline
${}^b$ assuming stellar luminosity from \cite{Delfosse2013}, semi major axis $a=0.156$~AU and Equation~(11) from \citet{Carone2015}.\newline
${}^c$ Using Equation~(23) from \citet{Carone2014}.\newline
\end{table*}

We investigate in this work the basic principles of stratosphere circulation for different possible circulation states on rocky, tidally locked habitable ExoEarths. Here, we no longer have to rely on hypothetical rocky exoplanets orbiting ultra-cool red dwarfs like in \cite{Carone2015}. Instead, we can directly apply our model to TRAPPIST-1b, TRAPPIST-1d, Proxima Centauri~b and GJ~667~C~f. When assuming Earth-like atmosphere composition and thermal forcing, these planets cover the entire possible circulation state regime with respect to orbital period identified by \cite{Carone2015} (Figure~\ref{fig: climate}).

Earth-like thermal forcing may indeed be a valid assumption for TRAPPIST-1d (with some caveats), Proxima Centauri-b and GJ~667~C~f, based on incident stellar irradiation (Table~\ref{tab: TR-param}). \citet{Wolf2017} conclude that TRAPPIST-1e may have Earth-like surface temperatures only with a higher content of greenhouse gases, whereas \citet{Turbet2017} state that TRAPPIST-1e should have liquid surface water  for many atmosphere scenarios - including Earth-like greenhouse. In any case, TRAPPIST-1b and -1c are receiving too much stellar irradiation for Earth-like thermal forcing. TRAPPIST-1f, -1g, and -1h receive, conversely,  too little stellar irradiation to have Earth-like surface temperatures with an Earth-like atmosphere composition. However, TRAPPIST-1f and -1g may still be habitable if they have a thick CO${}_2$-dominated atmosphere (>1~bar and >5~bar, respectively) \citep{Turbet2017,Kopparapu2013}. \citet{Luger2017} speculate that TRAPPIST-1-h may be habitable if it migrated from the inner region of the planetary system to its current location quickly enough to build an atmosphere that is significantly enriched in hydrogen via outgassing.

For the habitable TRAPPIST-1b scenario, we acknowledge that this simulation is for now a hypothetical scenario that would require a colder host star than TRAPPIST-1. More specifically, we find that a TRAPPIST-1b-equivalent habitable planet would require a star with effective temperature $T_{star,eff}=2095$~K, that is 340~K cooler than TRAPPIST-1, to reside at the very inner edge of the habitable zone. That is, if we fix the albedo to the highest possible value of 0.5 \citep{Yang2013,Yang2014,Boutle2017}, assume a stellar radius of $R_{star}=0.1 R_{Sun}$, appropriate for ultra-cool-M dwarfs and brown dwarfs (e.g. \citet{Gillon2017,Deleuil2008}), and calculate $I_0=\sigma T_{star,eff}^4\left(\frac{R_{star}}{a}\right)^2$, where $\sigma$ is the Stefan Boltzmann constant and $a$ the planet's semi major axis. We furthermore require $I_{net}=I_0(1-\alpha)=958$~W/m${}^2$ (see Table~\ref{tab: TR-param}). The same exercise yields for TRAPPIST-1c, which resides in the same circulation state regime, a hypothetical host star with $T_{star,eff}=2452$~K, only 100~K cooler than TRAPPIST-1. Due to the large number of cool dwarf stars and surveys like CARMENES that search for planetary systems around such stars \citep{Quirrenbach2010},  it is very much possible that potentially rocky planets with three days orbital periods or less will be found.

For TRAPPIST-1d, we assume Earth-like atmosphere composition and Earth-like greenhouse effect, which is realized in our simplified model by setting surface optical depth $\tau_{s}=0.68$. We further assume a net irradiation of $I_{net}=I_0(1-\alpha)=958$~W/m${}^2$, where $I_0$ is incident irradiation. These assumptions yield for TRAPPIST-1d an albedo $\alpha=0.38$. The habitability of TRAPPIST-1d is, however, contested as it may undergo a runaway greenhouse effect, if the planet is covered by oceans \citep{Gillon2017,Wolf2017,Turbet2017,Kopparapu2017}. Apparently, the 'parasol mechanism' by clouds at the substellar point, proposed by  \citet{Yang2014}  to avoid the runaway greenhouse effect at the inner edge of the habitable zone, is not effective for this planet. An understanding of the basic atmosphere dynamics explains why: The parasol mechanism was based on a circulation scenario for a tidally-locked planet with an orbital period of 60~days.  As noted in \cite{Yang2013,Yang2014,Carone2016}, the mechanism requires the strong unperturbed direct upwelling branch of the direct circulation cell at the substellar point. This is obviously the case for slowly rotating planets with orbital periods larger than 25~days, where direct circulation dominates (see e.g. Figure~\ref{fig: climate}). For a habitable TRAPPIST-1d scenario with an orbital period of 4.05~days, the strength of the direct circulation cell is diminished by at least one order of magnitude compared to the slowly rotating ExoEarths with $P_{orb} > 25$~days (see Figure~18 in \cite{Carone2016}). The relatively weak upwelling apparently does not produce the necessary cloud coverage to effectively shield the planet from stellar irradiation in the 3D models of \citet{Wolf2017,Kopparapu2017}; with the caveat that these models only captured one out of two possible circulation states.

 \cite{Abe2011,Zsom2013} have identified, however, another possibility to avoid a runaway greenhouse at the inner edge of the habitable zone: a reduced water content or a "desert planet" scenario. Calculations by \citet{Bolmont2017} also show that the planet may have lost some of its water content in the past, but that it can still retain some water, thus remaining habitable. In light of the uncertainties in circulation state and water content, we therefore still assume that TRAPPIST-1d may be habitable. In addition, the same statement holds for TRAPPIST-1d than for TRAPPIST-1b: Even if this particular planet may not be habitable, a similar planet around a cooler M dwarf star may be.

For Proxima Centauri~b, the incoming irradiation is lower than on Earth ($I_0=900$~W/m${}^2$). For this case, we assume an Earth-like albedo $\alpha=0.3$, but use a more efficient greenhouse effect with $\tau_s=0.94$. We thus assume the greenhouse efficiency found by \citet{Edson2011} for tidally locked ExoEarths in a dry planet set-up, where it was confirmed in \cite{Carone2016} that this set-up generally yields similar results in terms of large scale circulation dynamics compared to the less efficient more Earth-like greenhouse effect with  $\tau_s=0.62$.  Our prescription for Proxima Centauri~b is different from the set-up investigated by \cite{Boutle2017}, who chose a weaker greenhouse effect but with a global sea surface. Despite the weaker greenhouse, simulations by \citet{Boutle2017} still resulted in surface temperatures above the freezing point of water for a large fraction of the eternal day side in their 1:1 spin-orbit resonance case.

For GJ~667~C~f, we also use an increased greenhouse gas effect by assuming $\tau_s=0.94$ and Earth-like albedo. This set-up results in effective and surface temperatures that are about eight degrees colder than on Proxima Centauri b, placing this planet at the outer edge of its star's habitable zone. We select this non-transiting planet because it is one of the very few tidally locked planets  a) with an orbital period larger than 25~days, b) that is in the habitable zone of its star and c) has a mass small enough to ensure that the planet is indeed rocky and not a Mini-Neptune. For the same reason, planets with radii larger than 1.5 Earth radius were excluded. See e.g. \citet{Fulton2017} for the emerging picture of a radius and compositional gap between 1.5 and 2 Earth radii in the exoplanetary population.

Furthermore, the following assumptions were made to derive surface gravities: The surface gravity for all investigated planets can in principle be calculated via
\begin{equation}
g=GM_P/R_P^2, \label{eq: gravity}
\end{equation}
  if the planetary mass ($M_P$) and radius ($R_P$) are both known. For the TRAPPIST-1 planets, both properties are known, but their masses come with relatively large uncertainties compared to their radii. If the mass and radius ranges allow it, we assumed Earth-like density, that is, $\rho_P=\rho_{Earth}=5.51$~g/cm${}^3$. This is indeed the case for TRAPPIST -1d \citep{Gillon2017,Wang2017}. Recent updates on the masses of the TRAPPIST-1 planets by \cite{Wang2017} imply that the density of TRAPPIST-1b is smaller than the Earth. We take the updated mean mass estimate for this planet ($M_P=0.79 M_{Earth})$\footnote{We note for future updates on the masses of the TRAPPIST-1 planets that changes in the assumed planetary mass and thus in the estimated surface gravity by 50\% or less are not going to fundamentally change the circulation patterns discussed in this work.}.

 For Proxima Centauri b and GJ~667~C~f, there are no radius constraints as these planets are not transiting. Furthermore,"only" their minimum masses have been measured. For Proxima Centauri b, we select for now  the radius and mass estimated from the evolution models of \cite{Zuluaga2016}, which lead to Mars-like bulk composition (0.71~$\rho_{Earth}$). For GJ~667~C~f, we use the measured minimum mass $m \sin( i)$ as the real mass and assume Earth-like density.

Eventually, we derive surface gravity by inserting into Equation~\ref{eq: gravity} the planetary mass expressed as
 \begin{equation}
 M_P=\rho_P\frac{4}{3}\pi R^3_P\label{eq: M_P}.
 \end{equation}
Table~\ref{tab: TR-param} lists the planetary, atmospheric and stratosphere wind breaking parameters used in this work.

\section{Simulations}

\begin{figure*}
\includegraphics[width=0.45\textwidth]{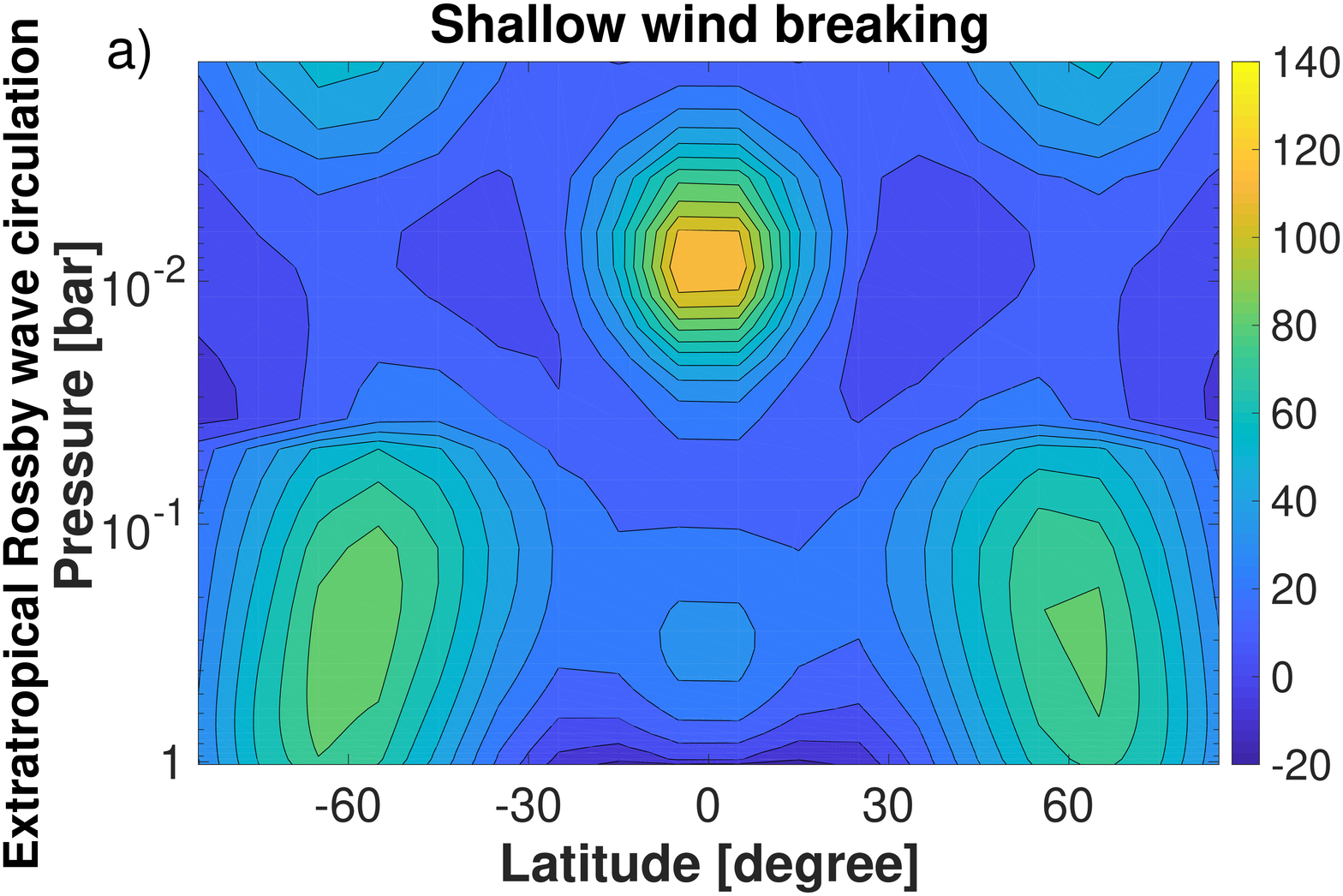}
\includegraphics[width=0.45\textwidth]{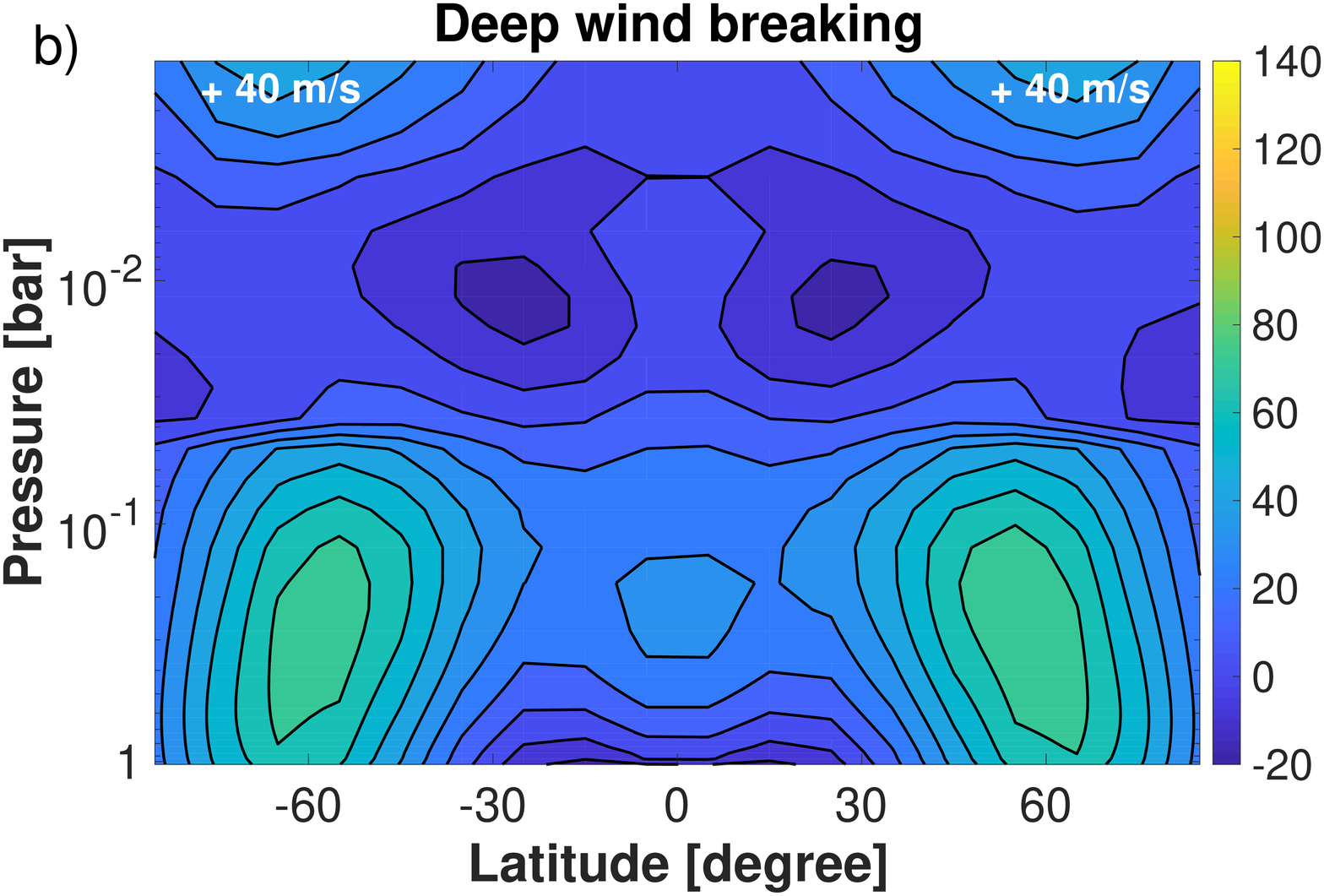}
\includegraphics[width=0.45\textwidth]{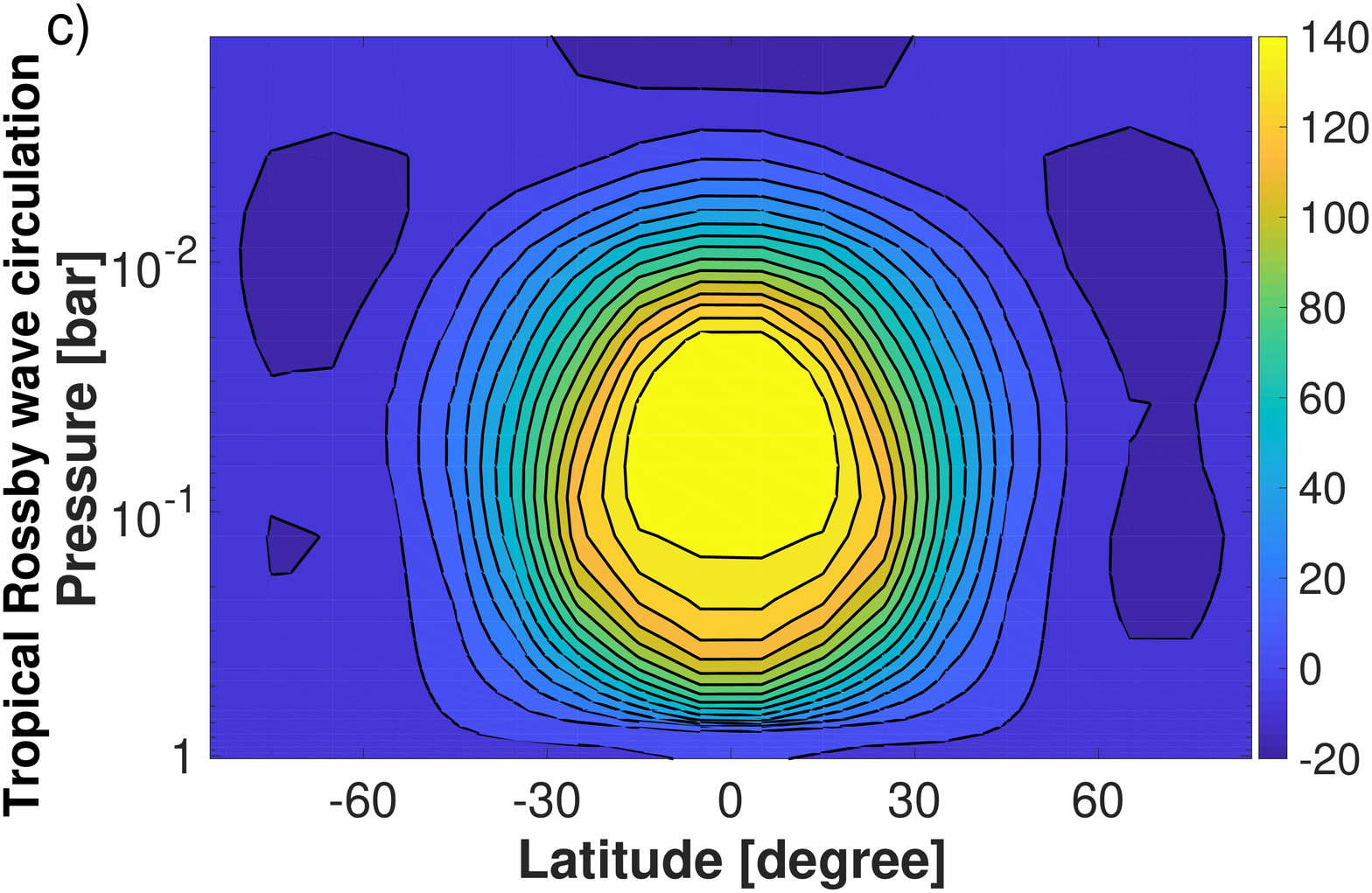}
\includegraphics[width=0.45\textwidth]{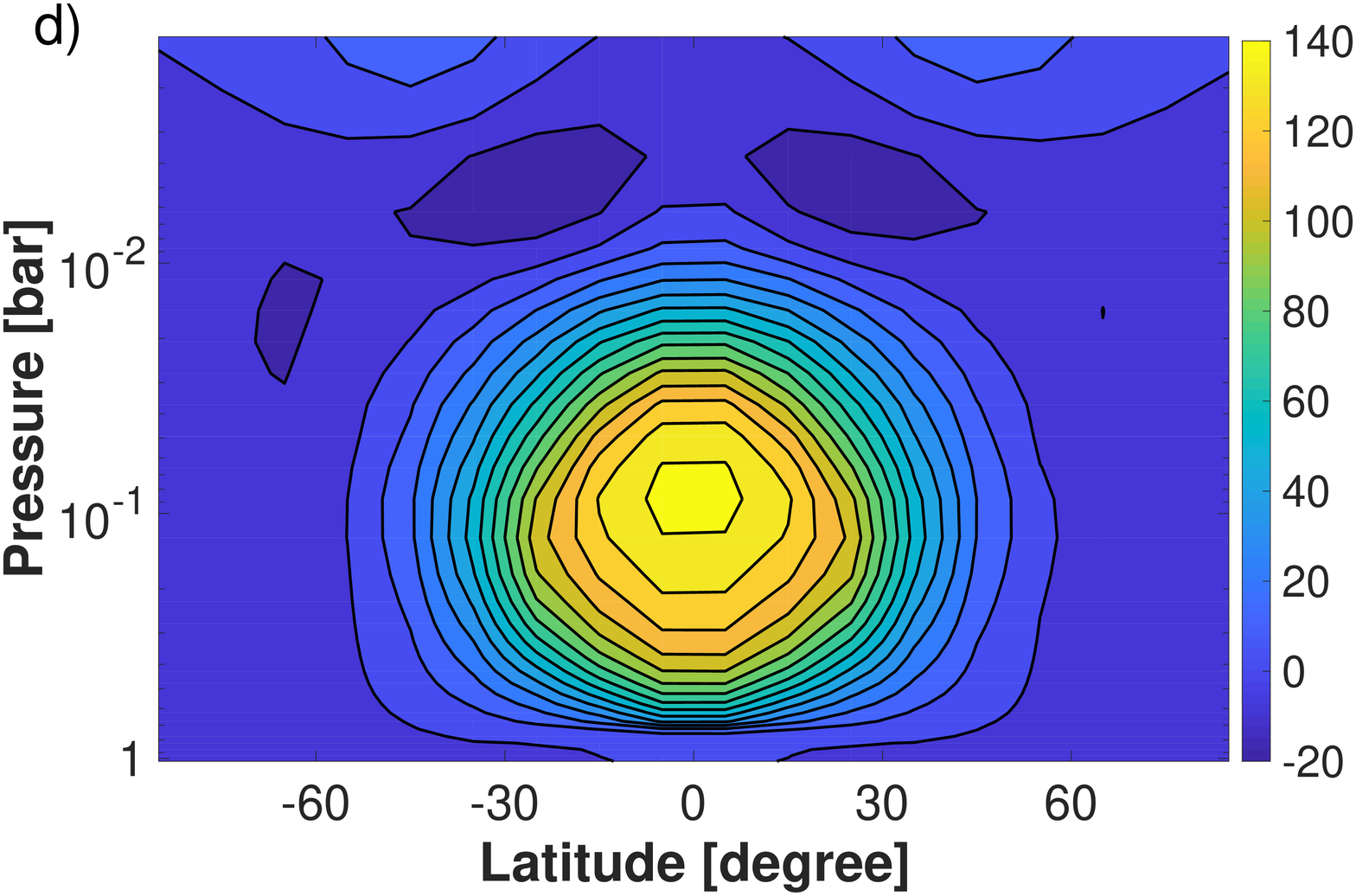}
\caption{The four faces of a habitable TRAPPIST-1b scenario: Zonally averaged zonal wind jets for identical thermal forcing but different circulation states and stratosphere wind breaking.  Contour level spacing is 10~m/s.}
\label{fig: TR-1b}
\end{figure*}

In the following, we investigate stratosphere wind systems and circulation for representative planets in each Rossby wave or circulation regime for different orbital periods as identified in \cite{Carone2015} (see also Figure~\ref{fig: climate}). These planets are TRAPPIST-1b, -1d, Proxima Centauri~b and GJ~667~C~f. For every circulation state, we investigate the deep and shallow stratosphere wind breaking scenario. For TRAPPIST-1b and -1d, we use different surface friction time scales to enforce two different circulation states for one and the same planet: one dominated by the tropical and one dominated by the extratropical Rossby wave. Thus, for every planet four to two different scenarios are investigated.

\subsection{Ultra-short orbital periods or the  TRAPPIST-1b scenario}

\begin{figure*}
\includegraphics[width=0.45\textwidth]{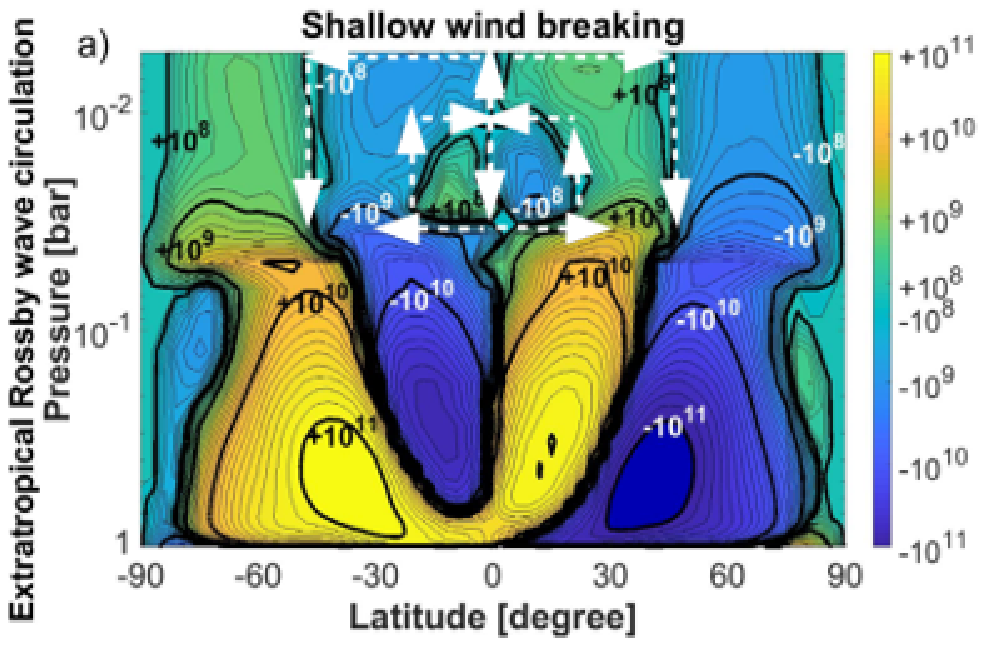}
\includegraphics[width=0.45\textwidth]{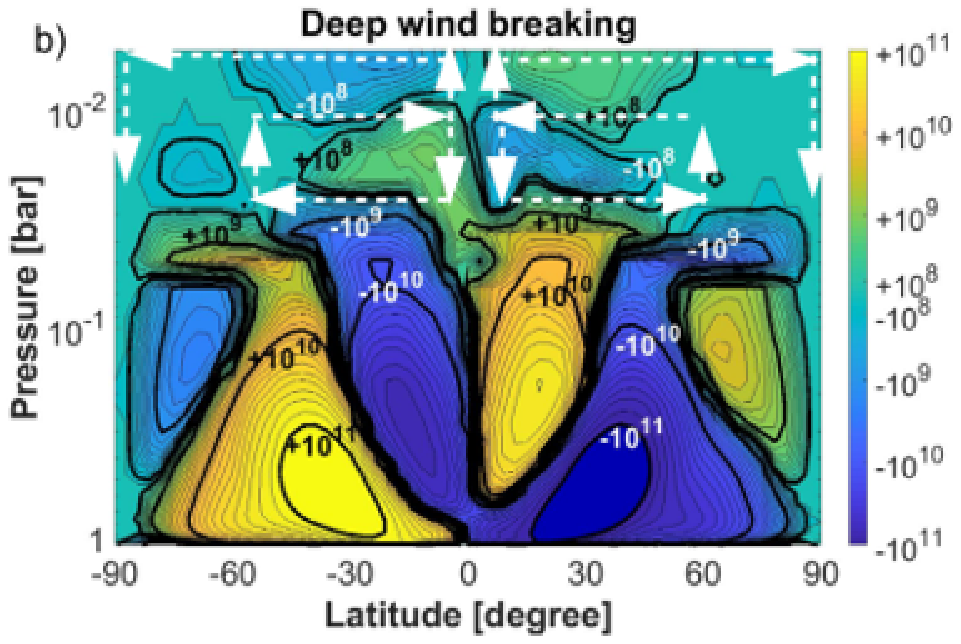}\\
\includegraphics[width=0.45\textwidth]{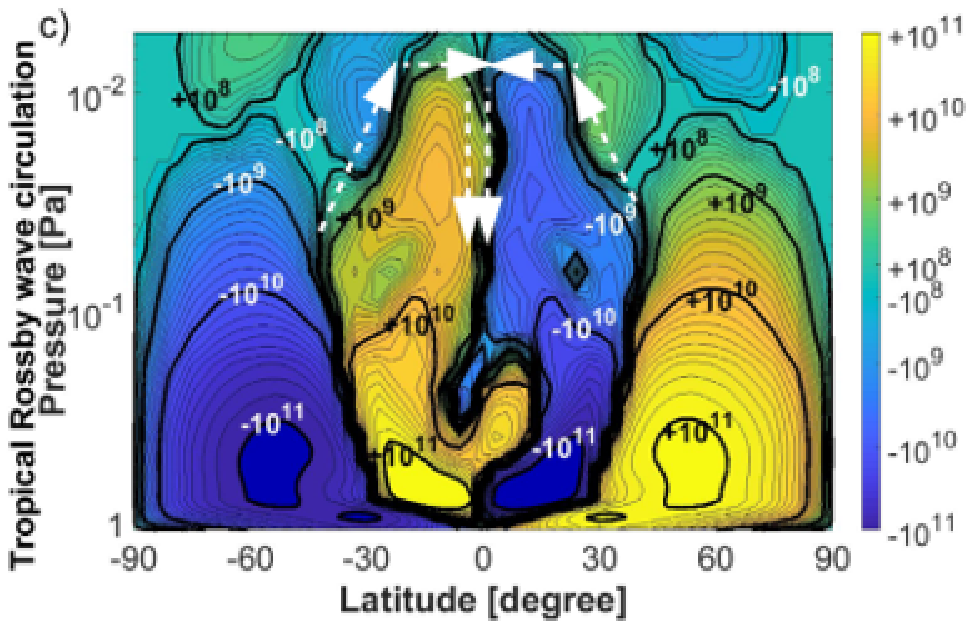}
\includegraphics[width=0.45\textwidth]{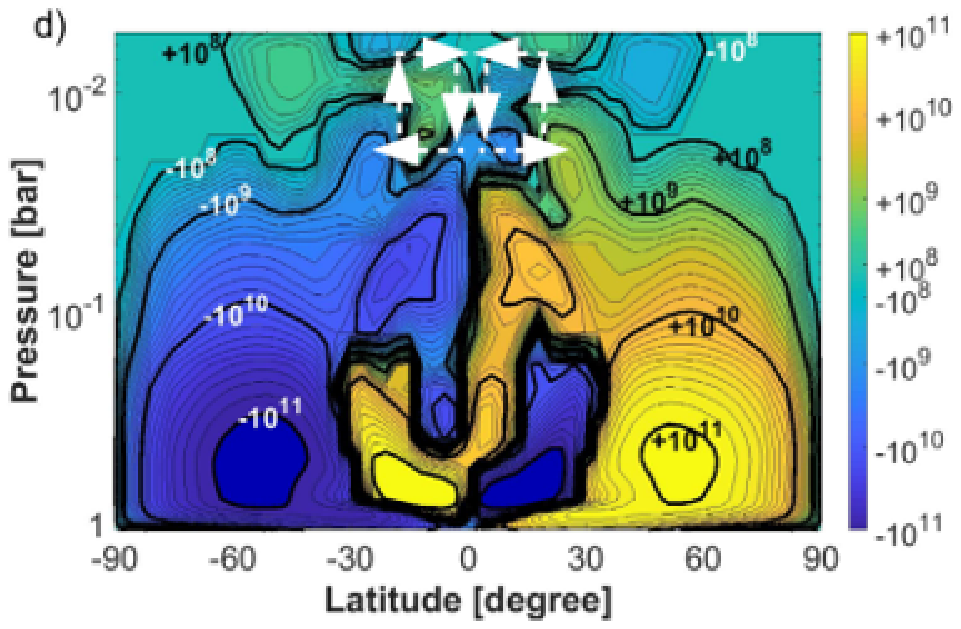}
\caption{The four faces of a habitable TRAPPIST-1b scenario:  Meridional stream function in kg/s  for identical thermal forcing but different circulation states and stratosphere wind breaking. Thin (thick) contour levels are arranged in steps of $10^{0.1}$ ($10^{1}$) between $10^{11}$ and $10^8$ ~kg/s and $-10^{11}$ and $- 10^8$ ~kg/s, respectively.  Dashed white arrows denote direction of circulation in the stratosphere.}
\label{fig: TR-1bcirc}
\end{figure*}

\begin{figure}
\includegraphics[width=0.49\textwidth]{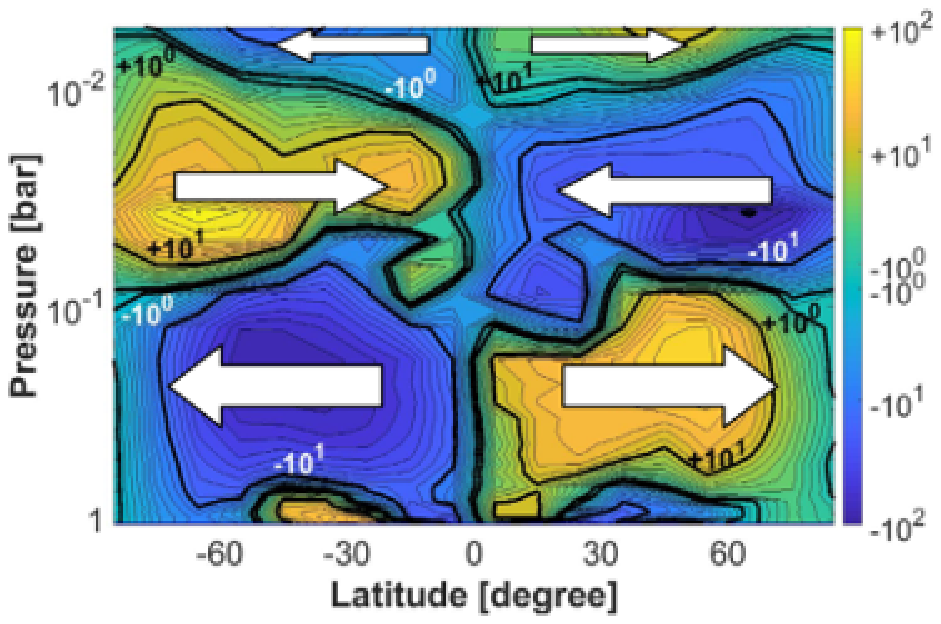}
\caption{TRAPPIST-1b: Eddy momentum flux in m${}^2$/s${}^2$ for extratropical Rossby circulation and deep stratosphere wind breaking. Thin (thick) contour levels are arranged in steps of $10^{0.1}$ ($10^{1}$) between $10^{2}$ and $10^0$~m${}^2$/s${}^2$ and $-10^{2}$ and $- 10^0$~m${}^2$/s${}^2$. White arrows denote direction of momentum transport.}
\label{fig: TR-1beddy}
\end{figure}

First, we study TRAPPIST-1b as a habitable ExoEarth scenario. This planet has an orbital period of 1.51~days (Table~\ref{tab: TR-param}).  As outlined in \cite{Carone2015} (see also Figure~\ref{fig: climate}), the circulation on such a planet is strongly determined by Rossby waves (tropical and extratropical). These lead to three possible 'banded' wind structures, that is, zonal wind jet regimes: The planet develops 1) two high latitude eastward wind jets 2) a strong eastward equatorial wind jet or 3) an equal mixture of both. In the first case, extratropical Rossby waves dominate. In the second case,  tropical Rossby waves dominate. In the third case, both waves are equally strong.

Generally, an extratropical Rossby wave circulation state is found to be much more clement than a tropical Rossby wave circulation state. In the tropical Rossby wave circulation state, strong equatorial wind jets in the troposphere severely disrupt heat transport from the day to the night side. One may even speak of an 'evil twin - good twin circulation scenario'.

Using \cite{Carone2016}, we set the surface friction time scale to  $\tau_{s,fric}=10$~days and  $\tau_{s,fric}=0.1$~days to reliably "toggle" between extratropical and tropical  Rossby wave circulation, respectively. The third mixed case, was not separately investigated. For each scenario, we further investigate two stratosphere wind breaking scenarios: the deep and shallow wind braking scenario as outlined in Section~\ref{sec: strato} (see also Table~\ref{tab: TR-param}).

Figure~\ref{fig: TR-1b} shows the zonal wind structure for the four investigated scenarios:
\begin{description}
\item{Scenario a)}  Extratropical Rossby wave circulation and shallow  wind breaking
\item{Scenario b)}  Extratropical Rossby wave circulation and deep wind breaking
\item{Scenario c)} Tropical Rossby wave circulation and shallow wind breaking
\item{Scenario d)} Tropical Rossby wave circulation and deep wind breaking.
\end{description}

Scenario a)  (Figure~\ref{fig: TR-1b}, top left) This simulation shows that while the extratropical Rossby wave dominates over the tropical Rossby wave in the troposphere  ($p>0.1$~bar), the situation reverses in the stratosphere ( $p<0.1$~bar). Consequently, we find an equatorial, superrotating stratospheric jet.  Circulation shows the development of  an 'Anti-Brewer-Dobson-circulation'  (Figure~\ref{fig: TR-1bcirc}, top left). The circulation is exactly reverse in direction compared to the Earth case (Figure~\ref{fig: Earth3}): instead of transporting air masses from the equator towards the poles, stratosphere circulation  rather confines them at the equator. Thus, chemical species produced in the stratosphere would in such a scenario tend to accumulate at the equator.

\begin{figure*}
\includegraphics[width=0.42\textwidth]{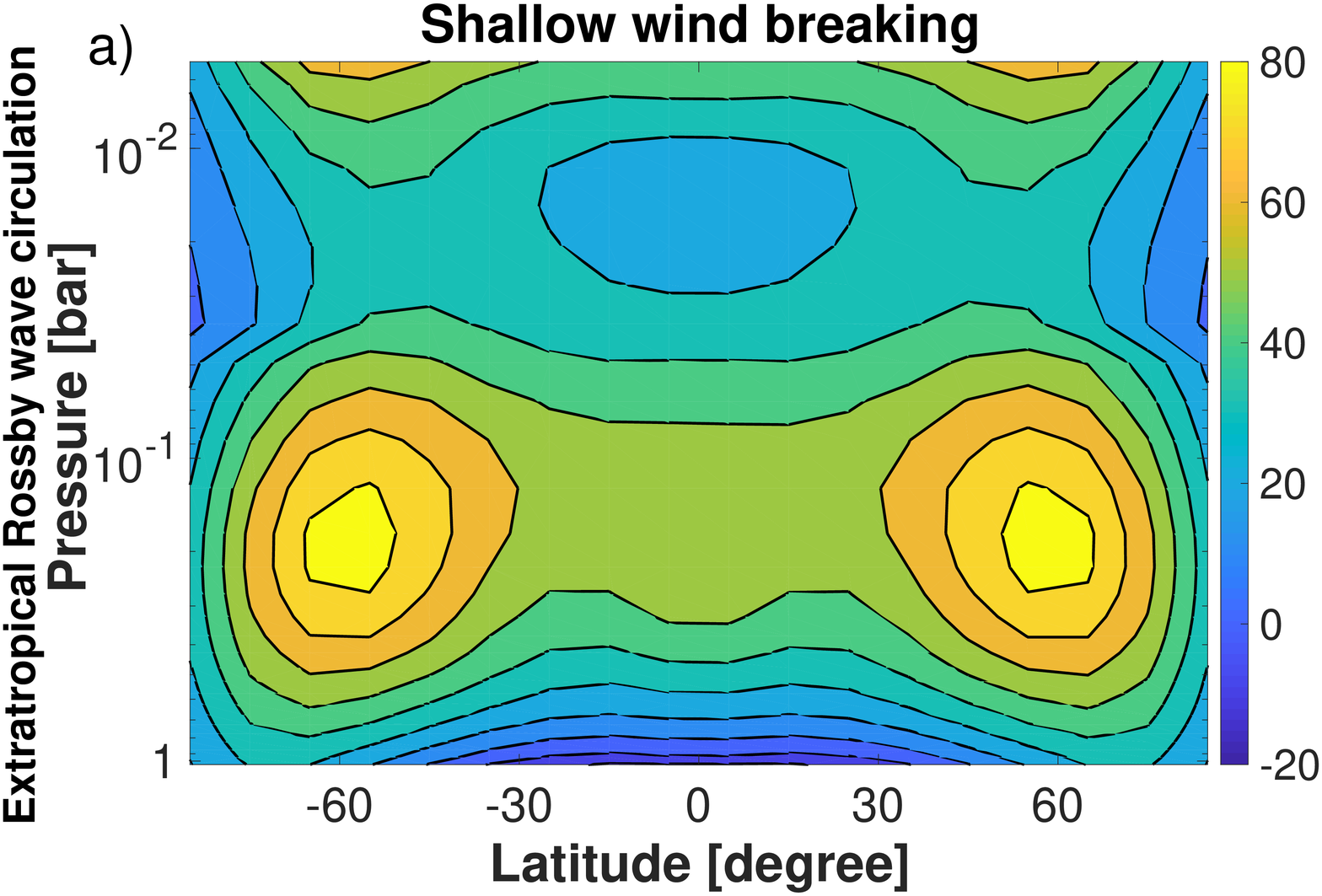}
\includegraphics[width=0.42\textwidth]{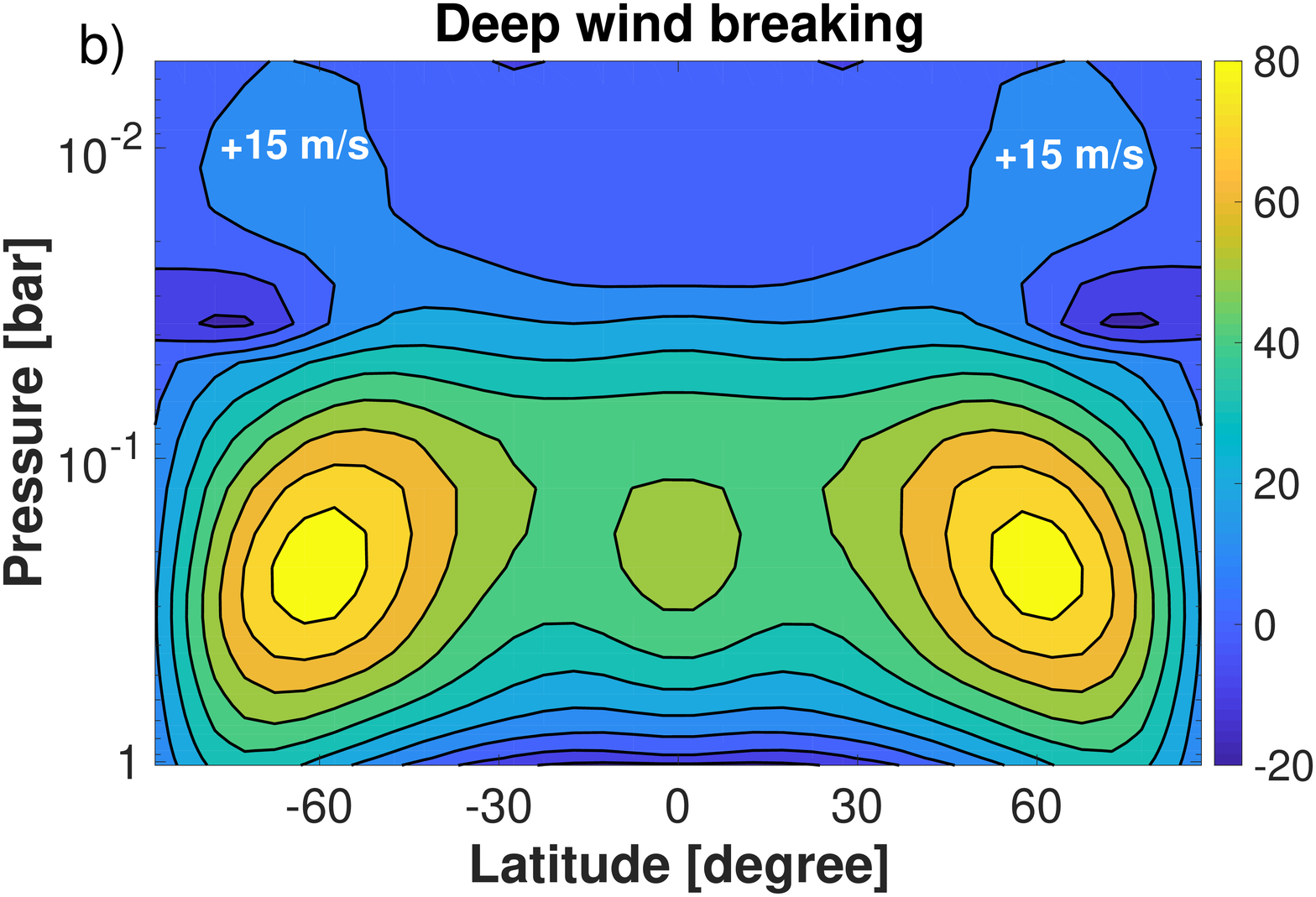}
\includegraphics[width=0.42\textwidth]{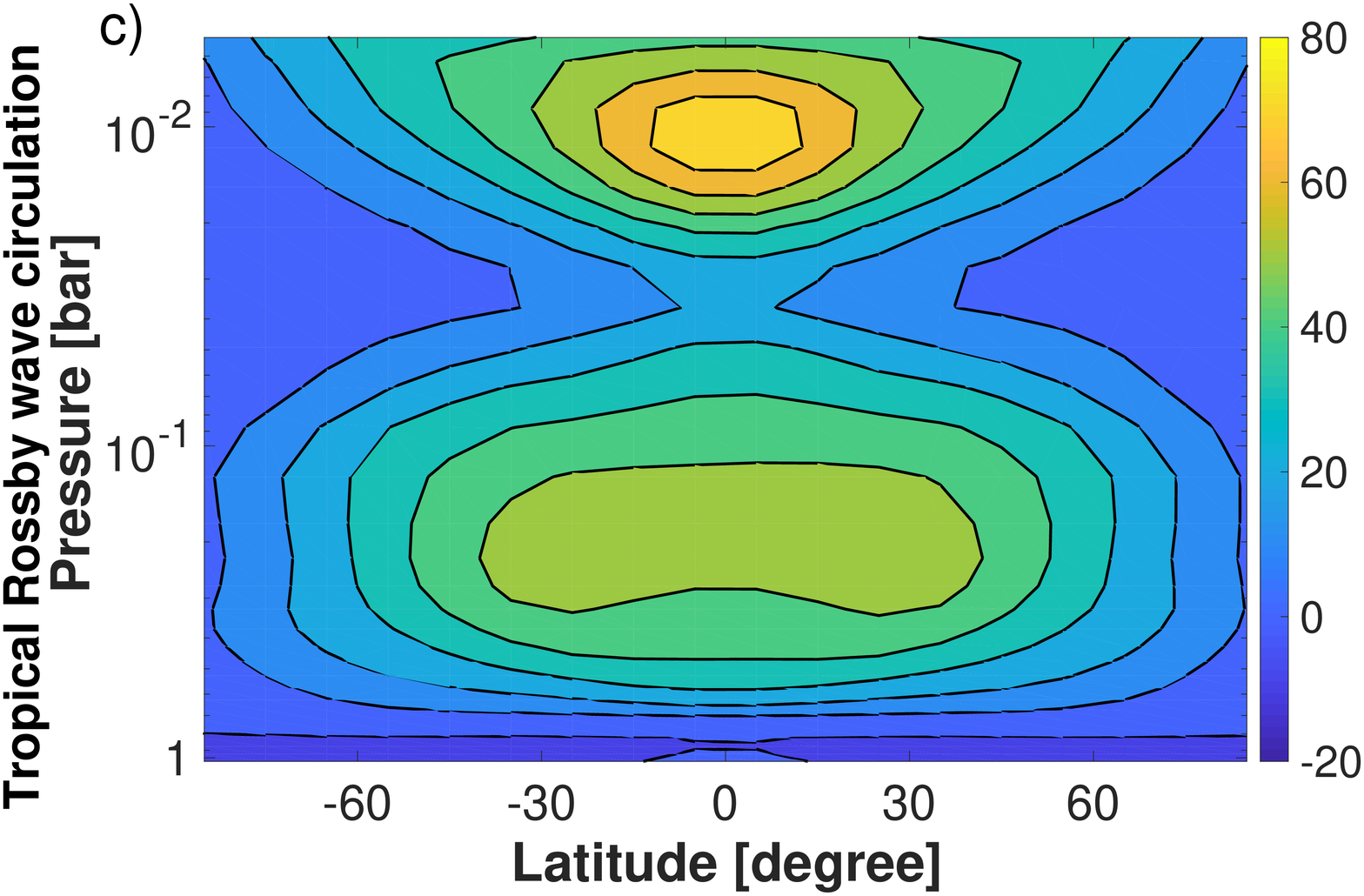}
\includegraphics[width=0.42\textwidth]{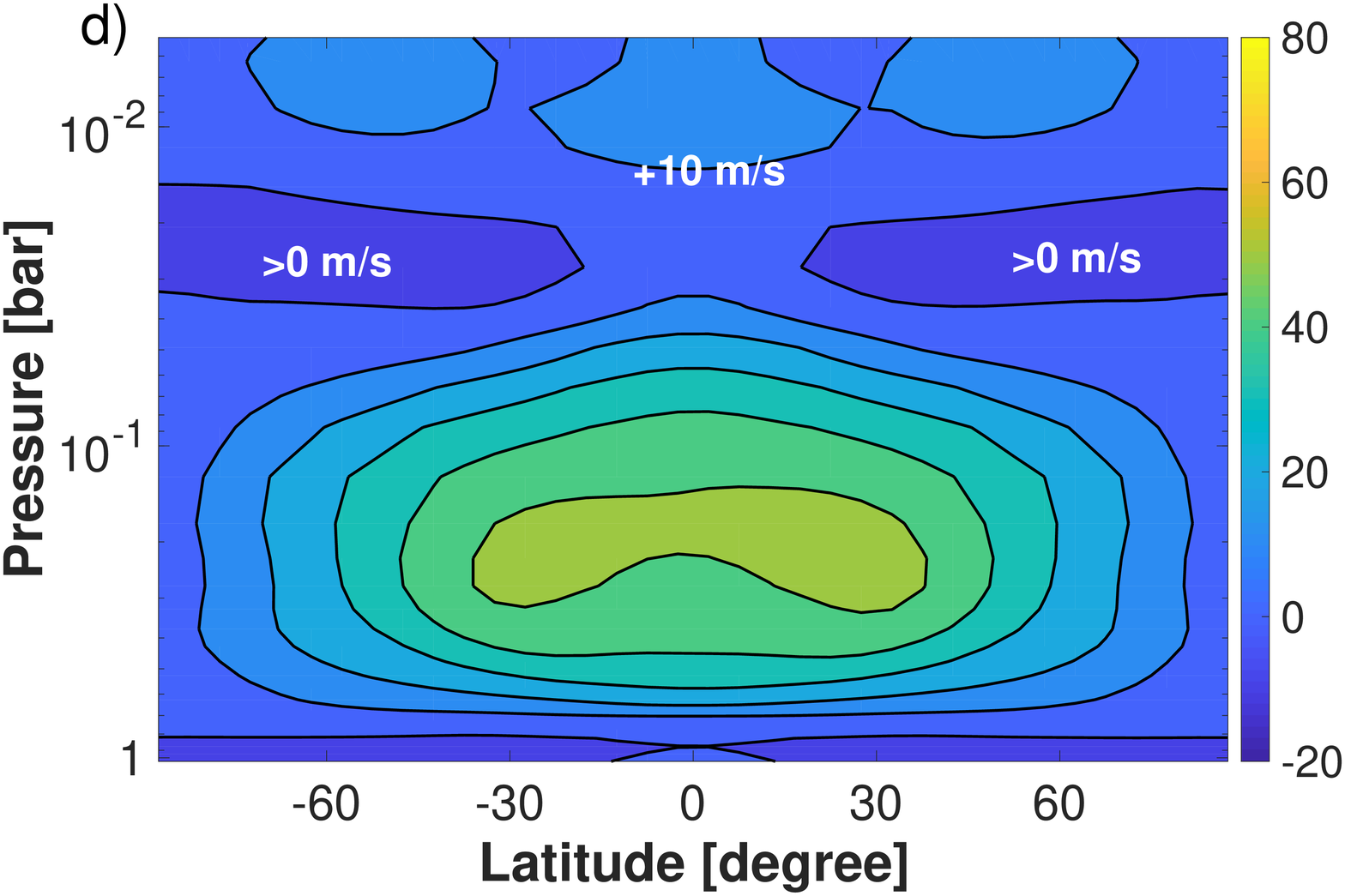}
\caption{The four faces of a habitable TRAPPIST-1d scenario: Zonally averaged zonal wind jets for identical thermal forcing but different circulation states and stratosphere wind breaking scenarios. Left panels show the deep stratosphere wind braking scenario. Contour level spacing is 10~m/s.}
\label{fig: TR-1d}
\end{figure*}

\begin{figure*}
\includegraphics[width=0.45\textwidth]{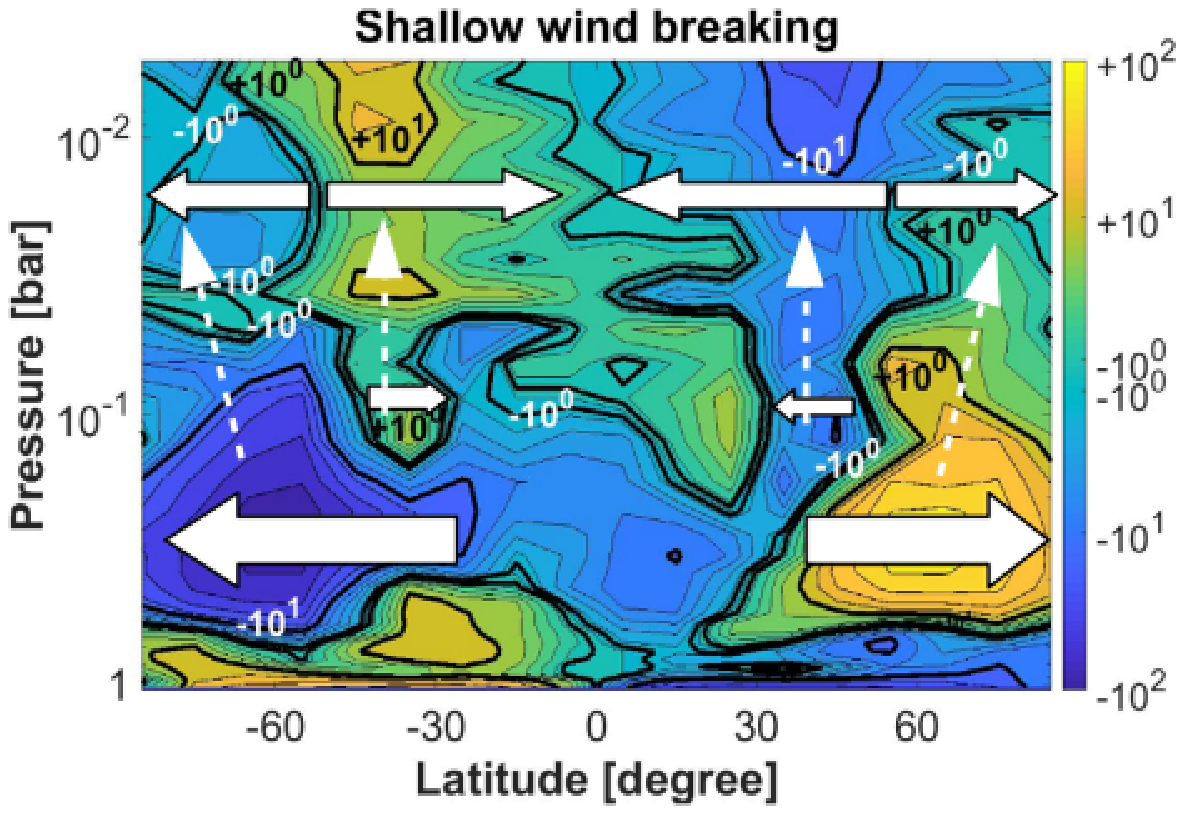}
\includegraphics[width=0.45\textwidth]{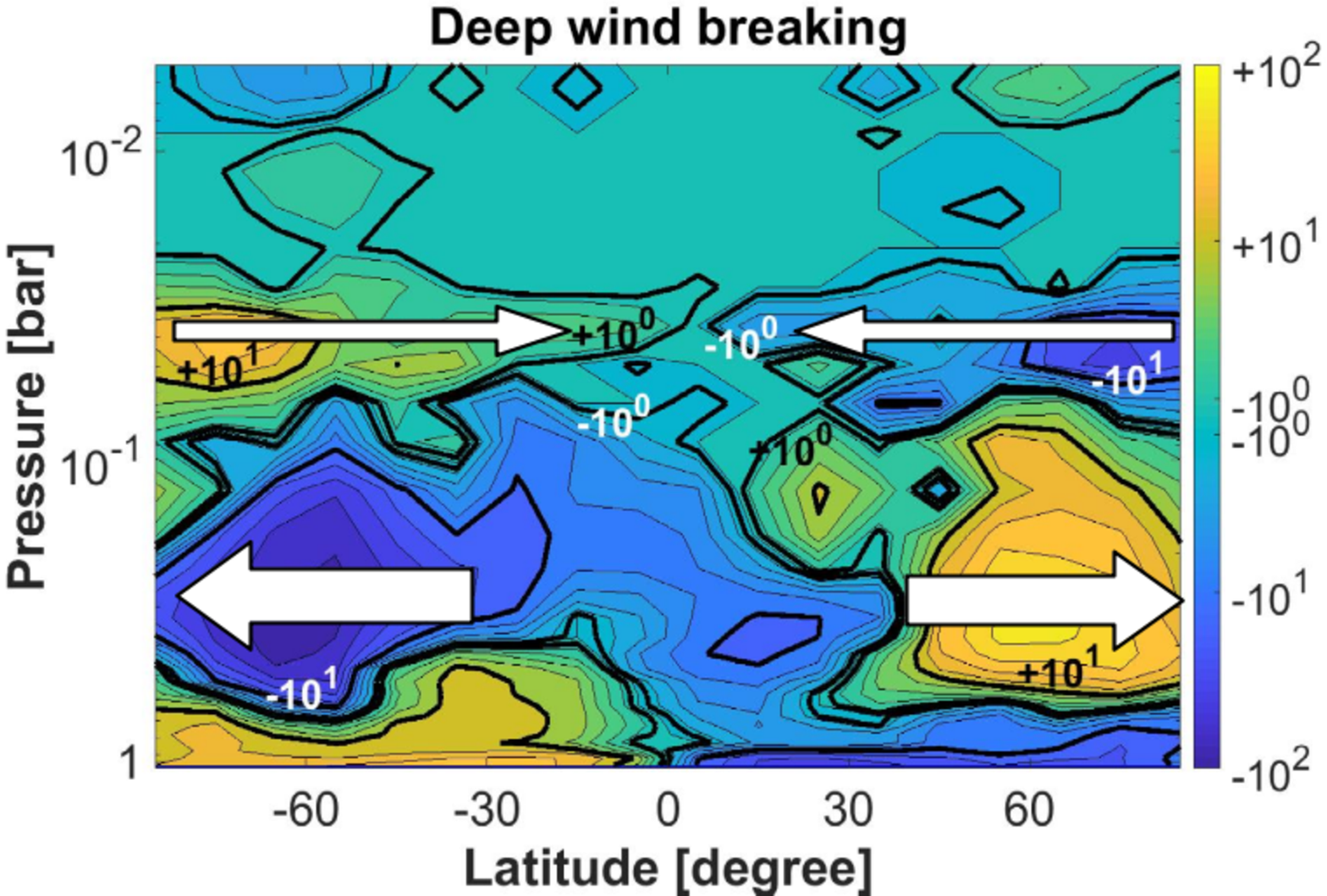}
\caption{Eddy momentum transport in TRAPPIST-1d's extratropical Rossby wave circulation scenario. Eddy momentum flux is in m${}^2$/s${}^2$. Thin (thick) contour levels are arranged in steps of $10^{0.2}$ ($10^{1}$) between $10^{2}$ and $10^0$~m${}^2$/s${}^2$ and $-10^{2}$ and $- 10^0$~m${}^2$/s${}^2$.   White arrows denote direction of momentum transport. Dashed arrows denote vertical propagation of waves.}
\label{fig: TR-1deddy}
\end{figure*}

Scenario b) The zonal wind map in  this simulation shows no equatorial stratospheric jet  (Figure~\ref{fig: TR-1b}, top right). Close inspection of the eddy momentum flux, however, shows that the tropical Rossby wave becomes here also dominant over the extratropical Rossby wave in the stratosphere. The action of the tropical wave is made apparent by the tendency to drive momentum equator-wards for  $p<0.1$~bar, whereas for  $p>0.1$~bar (troposphere), the extratropical Rossby wave drives momentum pole-wards (Figure~\ref{fig: TR-1beddy}). The absence of the stratosphere jet is thus entirely due to the effective friction mechanism of the deep wind breaking scenario. While the stratospheric equatorial wind jet is almost entirely suppressed, the same cannot be said for the associated Anti-Brewer-Dobson-circulation  (Figure~\ref{fig: TR-1bcirc}, top right). The tropical Rossby wave thus hampers distribution of air masses away from the equator even in this case, but not as severely as for shallow stratosphere wind breaking. Interestingly, a weak  Brewer-Dobson-like circulation with equator-to-pole-wards circulation is discernible above the Anti-Brewer-Dobson-circulation. This picture is confirmed by weak pole-wards eddy momentum transport visible at the top of the investigated atmosphere (Figure~\ref{fig: TR-1beddy}) and an associated weak polar  jet (Figure~\ref{fig: TR-1b}, top right). Some pole-wards transport is thus possible, if stratospheric winds are efficiently suppressed.

Scenario c) and Scenario d) Both cases are depicted in the bottom panels of Figures~\ref{fig: TR-1b} and \ref{fig: TR-1bcirc}. These two simulations show that the troposphere equatorial wind jet can extend far upwards into the stratosphere. Consequently, 'Anti-Brewer-Dobson-circulation' is very prominent and leaves no room for the Earth-like Brewer-Dobson-circulation. Therefore, again, species that are photo-chemically produced over the substellar point (such as potentially ozone) should accumulate over the equator and cannot be transported towards the polar regions. The main difference between shallow and deep stratosphere wind breaking scenario is that the jet extends farther upwards in the latter case.

In conclusion,  the 'evil circulation scenario' with a dominant tropical Rossby wave eventually wins out in the stratosphere for every investigated scenario. Planets with orbital periods less than three days will always suffer from Anti-Brewer-Dobson circulation that constrains air masses at the equator.  The specific circulation state and stratosphere wind breaking scenario 'only' determines to which degree stratosphere equator-to-pole-ward circulation is disrupted.

\subsection{Short orbital periods or the  TRAPPIST-1d scenario}

\begin{figure*}
\includegraphics[width=0.42\textwidth]{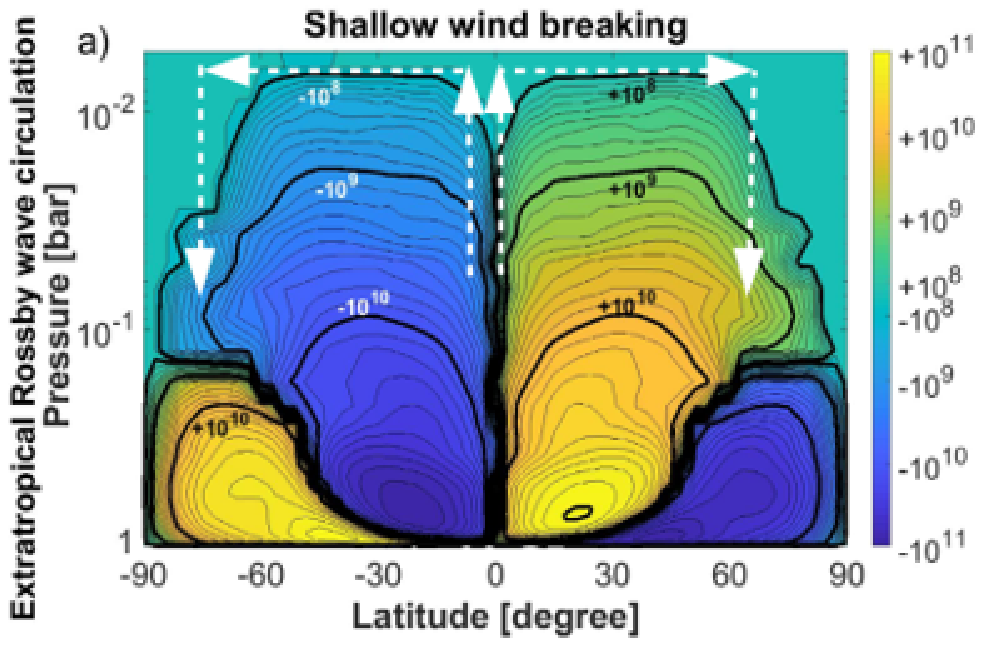}
\includegraphics[width=0.42\textwidth]{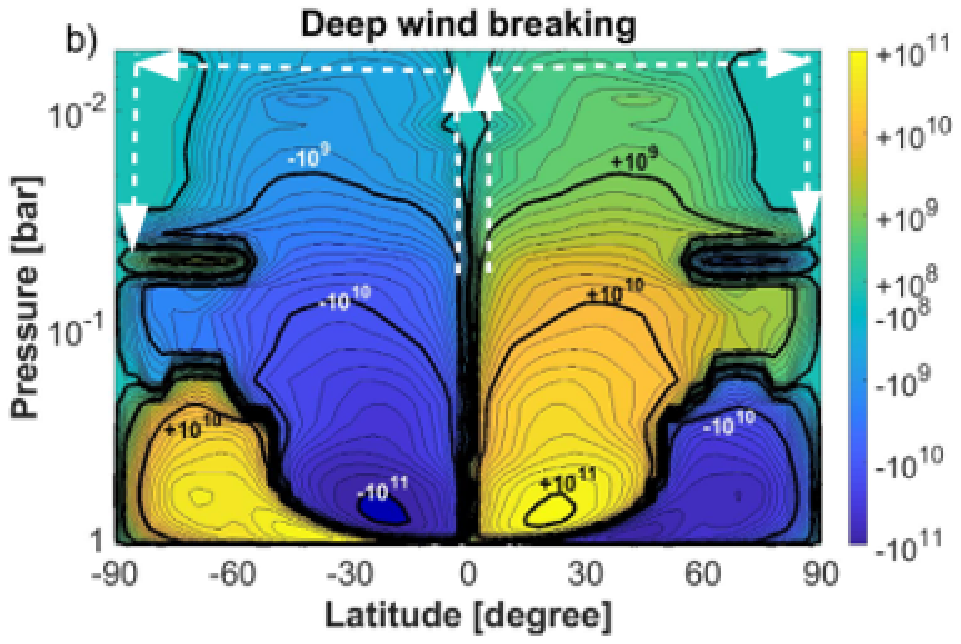}
\includegraphics[width=0.42\textwidth]{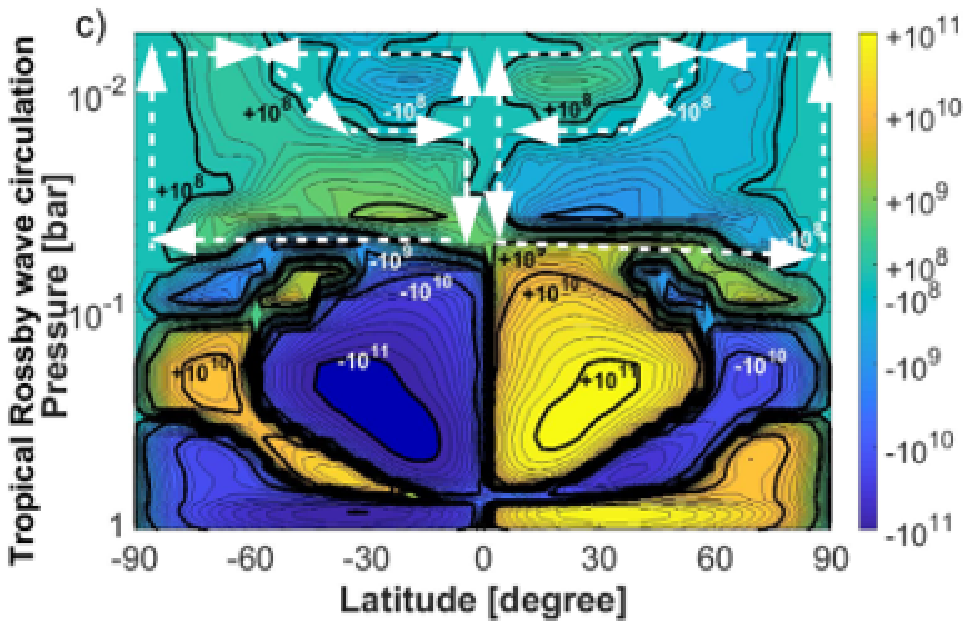}
\includegraphics[width=0.42\textwidth]{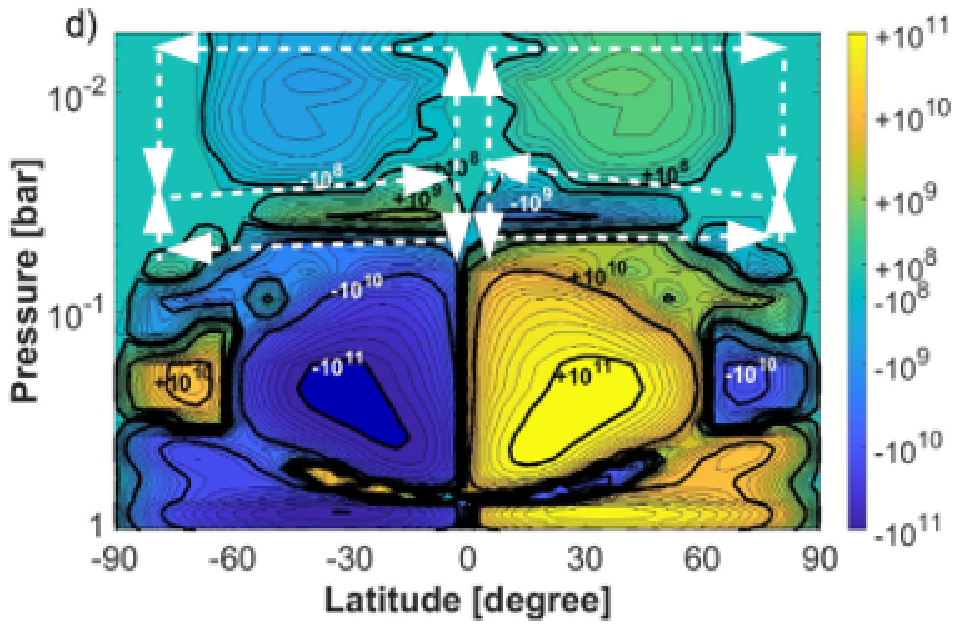}
\caption{The four faces of a habitable TRAPPIST-1d scenario:  Meridional stream function in kg/s  for identical thermal forcing but different circulation states and stratosphere wind breaking. Thin (thick) contour levels are arranged in steps of $10^{0.1}$ ($10^{1}$) between $10^{11}$ and $10^8$ ~kg/s and $-10^{11}$ and $- 10^8$ ~kg/s, respectively.  Dashed white arrows denote direction of circulation in the stratosphere.}
\label{fig: TR-1dcirc}
\end{figure*}
If Trappist-1d is habitable (see Section~\ref{sec: planet param}), its relatively short orbital period of 4.05~days will place the planet into the circulation region domain that is still strongly dominated by Rossby waves (Figure~\ref{fig: climate}).  TRAPPIST-1d - if tidally locked - can have two different circulation states for the same Earth-like thermal forcing. We "switch" again between those circulation states by changing the effective  surface friction time scale. We also investigate deep and shallow stratosphere wind braking scenarios (Figure~\ref{fig: TR-1d}).

Figure~\ref{fig: TR-1d} shows the zonal wind structure for the four investigated scenarios:
\begin{description}
\item{Scenario a)}  Extratropical Rossby wave circulation and shallow wind breaking
\item{Scenario b)}  Extratropical Rossby wave circulation and deep wind breaking
\item{Scenario c)} Tropical Rossby wave circulation and shallow wind breaking
\item{Scenario d)} Tropical Rossby wave circulation and deep wind breaking.
\end{description}

While the troposphere wind systems of TRAPPIST-1d resemble those of TRAPPIST-1b, the stratospheric wind jet systems are fundamentally different.

Scenario a) The extratropical Rossby wave dominates in the troposphere as evidenced by the two high latitude zonal wind jets for $p>0.1$~bar (Figure~\ref{fig: TR-1d}, top left). In the stratosphere, we find a second set of high latitude wind jets on top of the tropospheric ones.  These stratosphere wind jets superficially resemble Earth-like polar wind jets (Figure~\ref{fig: Earth}, top panel) that induce Brewer-Dobson circulation. Inspection of the eddy momentum transport, however, reveals that the underlying physics is not entirely Earth-like (Figure~\ref{fig: TR-1deddy}, left).

\begin{figure*}
\includegraphics[width=0.42\textwidth]{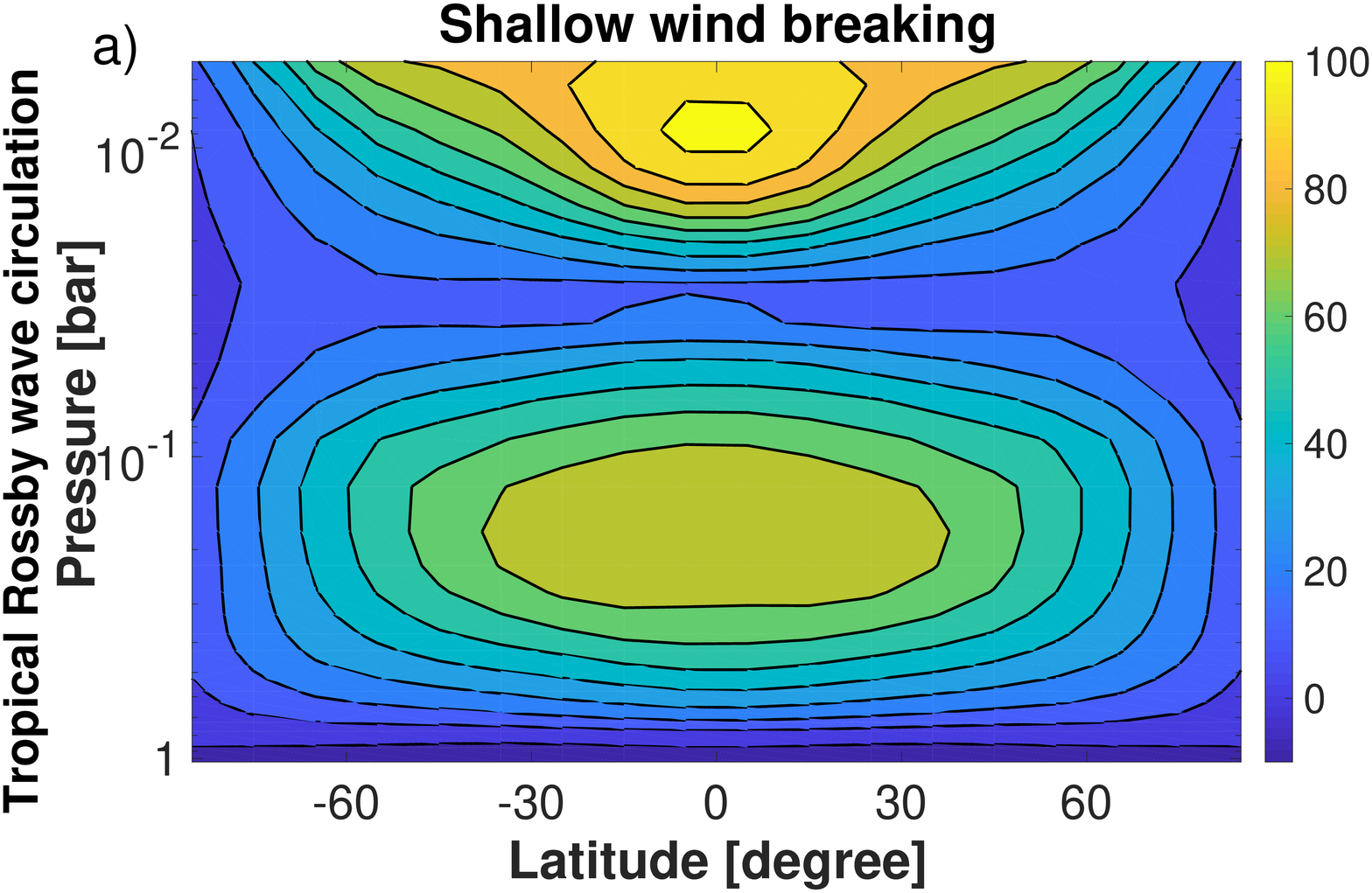}
\includegraphics[width=0.42\textwidth]{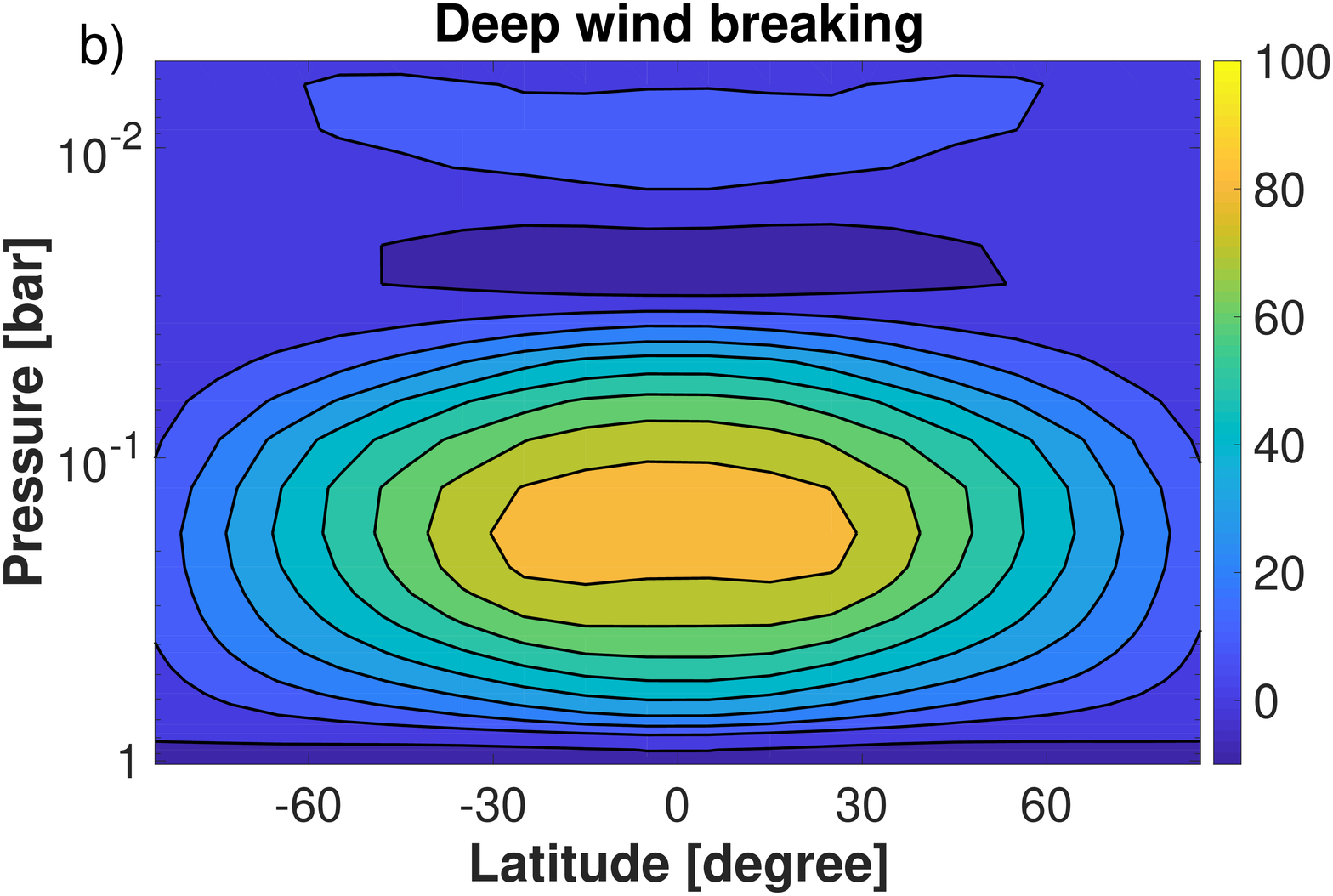}
\includegraphics[width=0.42\textwidth]{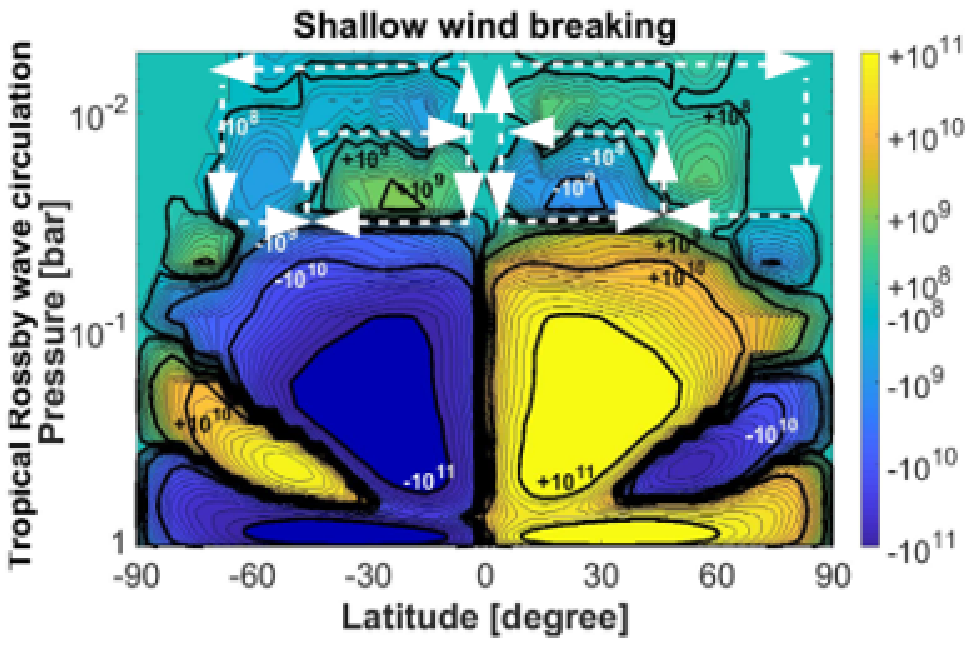}
\includegraphics[width=0.42\textwidth]{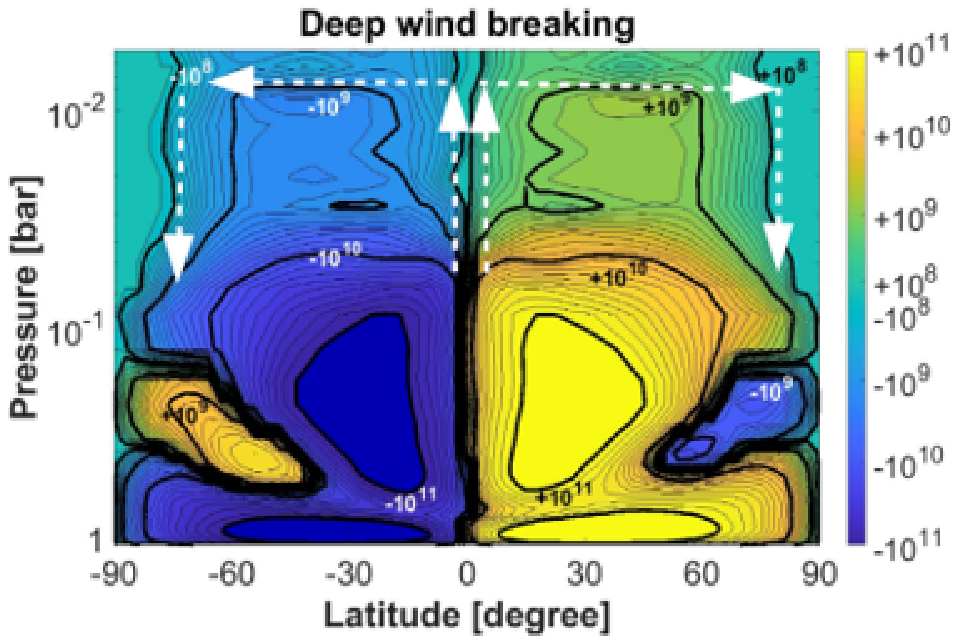}
\caption{The two faces of a habitable Proxima Centauri b scenario. Top panels: Zonally averaged zonal wind jets.  Contour level spacing is 10~m/s. Bottom panels: Meridional stream function in kg/s  for identical thermal forcing but different stratosphere wind breaking. Thin (thick) contour levels are arranged in steps of $10^{0.1}$ ($10^{1}$) between $10^{11}$ and $10^8$ ~kg/s and $-10^{11}$ and $- 10^8$ ~kg/s, respectively.  Dashed white arrows denote direction of circulation in the stratosphere. }
\label{fig: Prox}
\end{figure*}

The eddy momentum transport is in so far Earth-like as pole-wards momentum transport - associated with the extratropical Rossby wave - propagates vertically upward. See Figure~\ref{fig: Earth}, bottom panel, in the Earth case for comparison. The eddy momentum transport is, however, also un-Earth-like because also a standing tropical Rossby wave propagates vertically upwards and increases in strength. While the tropical Rossby wave never gains completely the upper hand in the stratosphere, as was the case for TRAPPIST-1b, it still manages to diminish the pole-wards momentum transport. The tropical Rossby wave, thus, probably also diminishes the equator-to-pole-wards circulation enabled by the extratropical Rossby wave (Figure~\ref{fig: TR-1dcirc}, top left).

 Scenario b) A weak  stratospheric high latitude wind jet system is present in the stratosphere. No clear pole-ward eddy momentum transport can be identified in the stratosphere (Figure~\ref{fig: TR-1deddy}, right). But neither is equator-ward momentum transport discernible. Therefore, both, tropical and extra tropical Rossby waves are efficiently suppressed by deep stratosphere wind breaking, which is unlike the Earth scenario (Figure~\ref{fig: Earth}, bottom panel). Since the Earth is rotating four times faster than TRAPPIST-1d  and since the strength of Rossby waves diminishes with slower rotation (see e.g. \citet{Showman2011,Holton,Carone2015}), it is no surprise that Rossby waves on TRAPPIST-1d are more easily suppressed than in the Earth case (Section~\ref{sec: strato}).

Conversely, the strength of  thermally driven circulation increases with slower rotation (see e.g. \cite{Carone2015,Held1980}).  Therefore, thermally driven circulation can assume the role on TRAPPIST-1d that the wave driven Brewer-Dobson-circulation has on Earth:  to transport air masses - and thus aerosols and photo-chemically produced species - from the equator towards higher latitudes. Indeed, stratospheric equator-to-pole-wards circulation is very dominant  (Figure~\ref{fig: TR-1dcirc}, bottom left). The circulation is even stronger than in Scenario a), where in principle the combination of thermally and wave driven circulation should lead to stronger circulation compared to only thermally driven circulation. We attribute this contradiction to the background presence of the tropical Rossby wave in Scenario a) that counteracts any pole-wards circulation.

Scenario c) This simulation shows more clearly the counteracting stratosphere circulation tendency by the tropical Rossby wave for the TRAPPIST-1d case. In the zonal wind picture, the tropical Rossby wave dominates in the troposphere and drives an equatorial wind jet there (Figure~\ref{fig: TR-1d}, bottom left). In the circulation picture, Anti-Brewer-Dobson-circulation is clearly discernible (Figure~\ref{fig: TR-1dcirc}, bottom left), but does not completely supersede the thermally driven pole-wards circulation, as was the case for TRAPPIST-1b (see previous Subsection). Another difference to the circulation scenario c) of TRAPPIST-1b is that the equatorial wind jet in the troposphere does not transcend into the stratosphere (Figure~\ref{fig: TR-1d}, bottom panels). Instead,  a second stratospheric equatorial wind jet develops on top that may still disturb pole-ward transport (Figure~\ref{fig: TR-1d}, bottom left).

Scenario d) For deep stratospheric wind breaking, the formation of a second stratospheric equatorial wind jet on top of the tropospheric one is clearly inhibited (Figure~\ref{fig: TR-1d}, bottom right). The Anti-Brewer-Dobson circulation is even less capable to disrupt the thermally driven equator-to-polewards stratosphere circulation (Figure~\ref{fig: TR-1dcirc}, bottom right).

Generally, we can conclude that the slower orbital period of TRAPPIST-1d compared to TRAPPIST-1b makes a global transport of aerosols and photo-chemical produced compounds more likely. In the extratropical Rossby wave circulation state, the tropical Rossby wave is efficiently prevented from creating havoc in the stratosphere, with and without deep stratosphere wind breaking. Thermally driven circulation establishes efficient equator-to-polewards transport. Since the tropical  Rossby wave is weaker, an equatorial stratospheric wind jet and Anti-Brewer-Dobson circulation that could hamper day side-wide transport are also weaker; even in the worst case scenario, when the troposphere is in the tropical Rossby wave circulation state and when stratosphere wind breaking is inefficient.  

\subsection{Weak superrotation or the  Proxima Centauri b scenario}

\begin{figure*}
\includegraphics[width=0.45\textwidth]{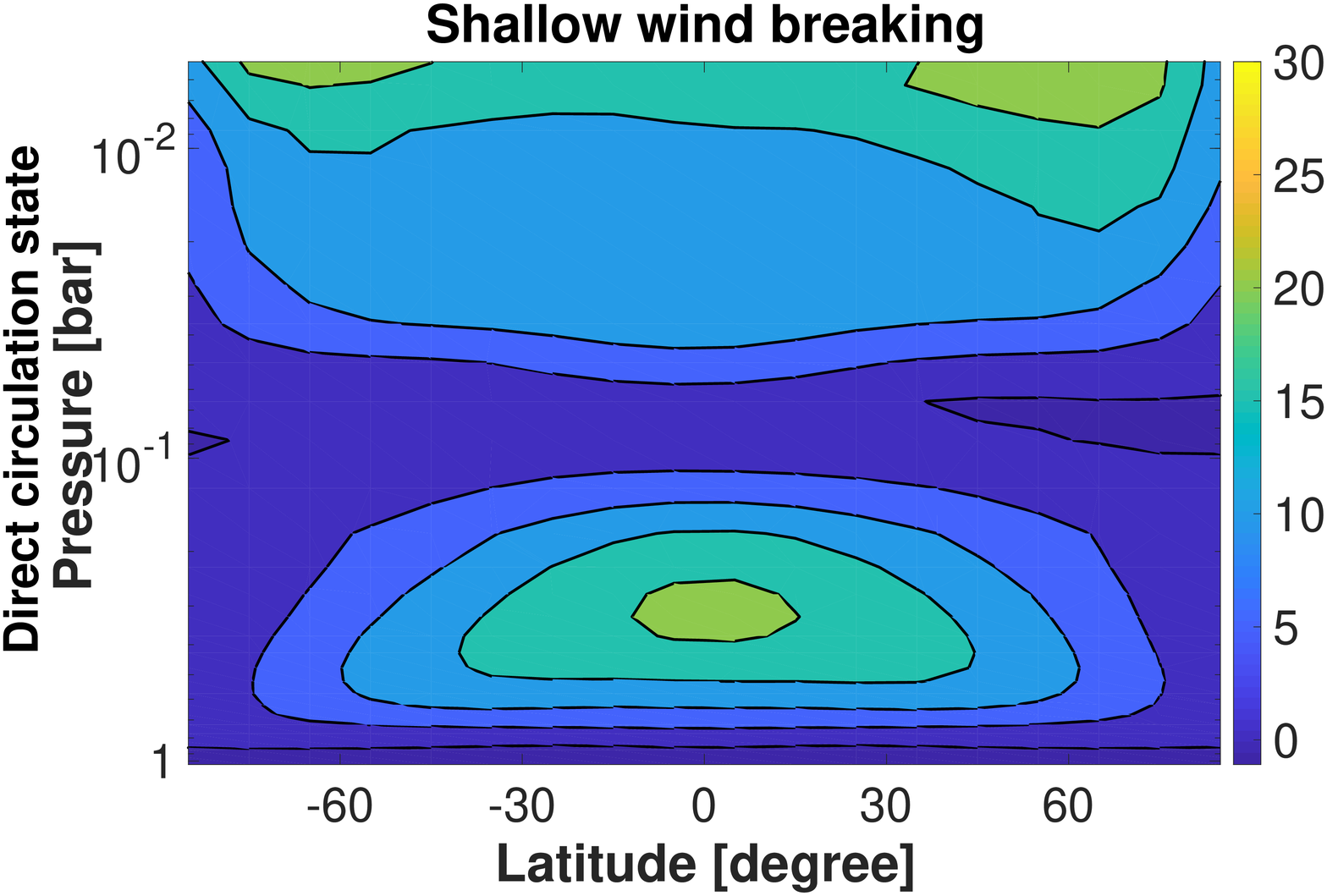}
\includegraphics[width=0.45\textwidth]{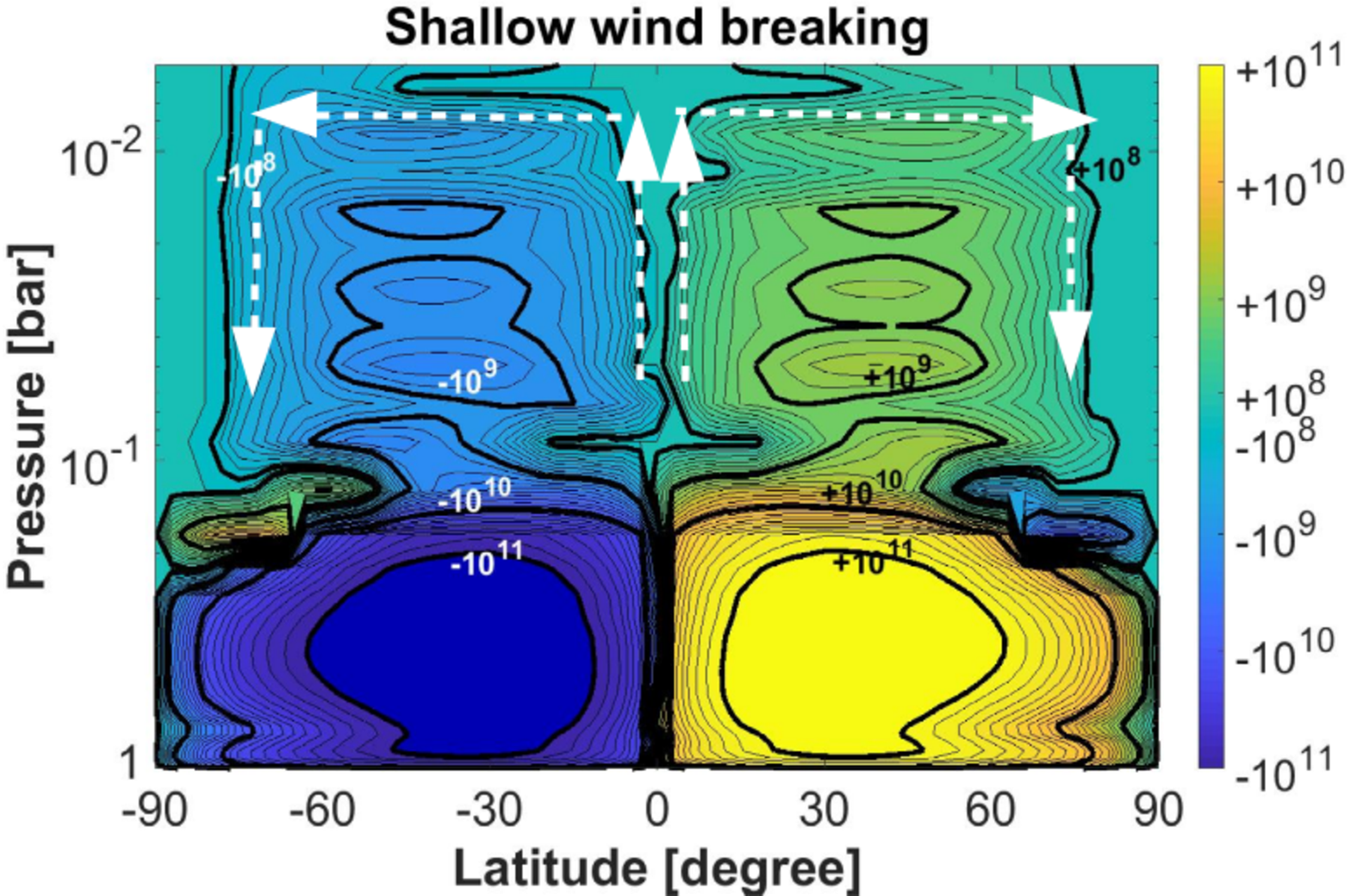}
\caption{The habitable GJ 667 C e scenario.  Left: Zonally averaged zonal winds. Contour level spacing is 5~m/s.  Right: Meridional stream function in kg/s. Thin (thick) contour levels are arranged in steps of $10^{0.1}$ ($10^{1}$) between $10^{11}$ and $10^8$ ~kg/s and $-10^{11}$ and $- 10^8$ ~kg/s, respectively.  Dashed white arrows denote direction of circulation in the stratosphere. Both properties are shown for shallow wind breaking.}
\label{fig: TR1h}
\end{figure*}

If TRAPPIST-1e to 1-h, LHS~1140~b, GJ~667 C c and Proxima Centauri b are habitable tidally locked ExoEarths, they will lie in the weak superrotation regime (Figure~\ref{fig: climate}). They will thus be very similar in circulation state, assuming a habitable atmosphere. We choose Proxima Centauri b as the prototype case for this circulation state region. 

Because standing extratropical Rossby waves cannot form for orbital periods greater than six days (assuming Earth size), only one circulation state scenario is possible on Proxima Centauri b. Thus, we don't assume extreme values of surface friction to switch between different circulation states as in the previous sections. Instead, we adopt for this planet the nominal surface friction time scale of $\tau_{s,fric}=1~$days. Two scenarios are considered in the following:
\begin{description}
\item{Scenario a)} Weak tropical Rossby wave circulation with shallow wind breaking
\item{Scenario b)} Weak tropical Rossby wave circulation and deep stratosphere wind breaking scenario.
\end{description}

Scenario a) Although Proxima Centauri b has a relatively weak tropical Rossby wave compared to TRAPPIST-1b and 1d, it is still strong enough to induce a strong equatorial wind jet in the stratosphere  with Anti-Brewer-Dobson-circulation  (Figure~\ref{fig: Prox}, left panels).

Scenario b) Here, the tropical Rossby wave is efficiently damped in the stratosphere: No equatorial jet and no Anti-Brewer-Dobson circulation can be found in the stratosphere (Figure~\ref{fig: Prox}, right panels).

Proxima Centauri b appears thus to be more prone to the adverse effects of the tropical Rossby wave than the habitable TRAPPIST-1d case (see previous Subsection) - even though the disruptive tropical Rossby wave is weaker. This counter-intuitive result can be explained when we consider that the extratropical Rossby wave is missing for Proxima Centauti b: Only deep, efficient stratosphere wind breaking  can 'save' the planet from the adverse effects of the tropical Rossby wave. In this case, the strong thermally driven circulation can take over the role that the wave driven Brewer-Dobson-circulation has on Earth: to transport aerosols and photo-chemically produced species from the equator to the poles.

If the tropical Rossby wave is not efficiently damped in the stratosphere, the formation of an equatorial 'transport belt' instead of a day side-wide transport regime is a distinct possibility, like for TRAPPIST-1b.

\subsection{Direct circulation dominance or the GJ 667 C f scenario}

We take GJ 667 C f as the prototype for a circulation state entirely dominated by strong direct circulation (Figure~\ref{fig: climate}). Since only very weak zonal wind tendencies can be found, we display only one case: Direct circulation with weak  stratosphere wind breaking

 Figure~\ref{fig: TR1h} (left) shows indeed that a tidally locked habitable ExoEarth with a 39~days orbital period has no strong eastward wind tendencies; neither in the troposphere nor in the stratosphere. Consequently, direct circulation is very efficient in  the troposphere and stratosphere and can transport material in the stratosphere over the whole day side. Indeed, the wind flow on such planets is not banded but mainly radial (see Figure~\ref{fig: climate}) and flows on top of the troposphere from the substellar point towards polar regions and the day side-night side terminators. 
  
We speculate that, for tidally locked planets with orbital periods greater than 25~days, the amount of ozone produced on the day side may be distributed over the whole day side and may even be depleted by vigorous transport away from the production region towards the night side by direct circulation. This picture is confirmed by \citet{Proedrou2016} who modeled a tidally locked Earth with an orbital period of 365~days: Their Figure~4 shows indeed a relatively uniform ozone distribution over the whole day side in the stratosphere. Even higher up, in the mesosphere, transport time scales are faster than chemical reaction time scales, leading  to depletion of ozone at the day side and accumulation of ozone at the night side.
 
On such a planet, mesospheric ozone that is streaming towards the night side could maybe be detected via transmission spectroscopy that probes the limbs of the planetary atmosphere \citep{Barstow2016}. Stratospheric ozone that stays on and is distributed over the whole day side could rather be detected by day side thermal emissions, as proposed by \citet{Kreidberg2016}.
 
However, future work is warranted to better connect circulation, photo-chemistry and observations and to investigate if the results found by \citet{Proedrou2016} still hold for planets like GJ 667 C f with shorter orbital periods and thus less vigorous direct circulation. Furthermore, future work needs to take into account that the host stars of such planets are much redder than the Sun and are frequently flaring. 

\section{Summary and discussion}
\label{sec: summary}
Stratosphere circulation on rocky exoplanets is determined by planetary waves that form in the underlying troposphere and propagate upwards. Thus, the state of stratosphere circulation may reflect the state of troposphere circulation, even if the latter cannot be directly observed. For tidally locked ExoEarths, in contrast to Earth, the standing tropical Rossby wave is of supreme importance and can even suppress day side-wide stratosphere circulation: Tropical Rossby waves induce a strong equatorial superrotating jet in the stratosphere, with wind speeds of about 100~m/s, and Anti-Brewer-Dobson-circulation that confines air masses at the equator.  Both effects disrupt equator-to-pole-wards air mass transport, which is ensured on Earth by the normal Brewer-Dobson-circulation, driven by extratropical Rossby waves.

All ExoEarths with orbital periods shorter than 25 days are in danger from these disruptions. That is, stratosphere circulation on Proxima Centauri b, LHS~1440 b, TRAPPIST-1b -1g may be confined to an equatorial 'transport belt'.

More specifically, we found that tidally locked ExoEarths with very short orbital periods ($P_{orb}\leq 3$~days), like TRAPPIST-1b, are very unlikely to have a equator-to-pole-wards stratosphere circulation: Tropical Rossby waves always dominate in the stratosphere, regardless if the troposphere circulation is dominated by the tropical Rossby wave or by its antagonist, the extratropical Rossby wave. Thus, stratosphere circulation would be severely hampered by the associated Anti-Brewer-Dobson-circulation. Aerosols and photo-chemically produced substances could not be efficiently redistributed over the day side towards higher latitude. Instead, they may accumulate over the equator. While TRAPPIST-1b itself is not habitable, the results are directly transferable to exoplanets around stars or brown dwarfs that are even colder than TRAPPIST-1 (see Section~\ref{sec: planet param}). 

The larger the orbital period, the less problematic the tropical Rossby wave becomes - if it can be counterbalanced by an extratropical Rossby wave or if it's damped by stratosphere wind breaking. Therefore, TRAPPIST-1d with an orbital period of 4.05~days exhibits efficient equator-to-pole-wards transport in the troposphere and the stratosphere - as long as the planet's troposphere is dominated by the extratropical Rossby wave. This is the only circulation state modeled by \cite{Wolf2017,Kopparapu2017}. If the troposphere of TRAPPIST-1d is dominated by the tropical Rossby wave instead, the tropical Rossby wave also dominates in the stratosphere - but can be dampened there by deep stratosphere wind breaking. Then, again thermally driven circulation can ensure efficient equator-to-pole-wards transport in the stratosphere. 

The standing extratropical Rossby wave can only form for  short orbital periods ($P_{orb}\leq$~6 days for 1 Earth radii planets), but the standing tropical Rossby wave forms for intermediate orbital periods $6 \leq P_{orb} \leq 25$~days. Proxima Centauri~b, for example, lies in this orbital period regime  (Figure~\ref{fig: climate}). Without the extratropical Rossby wave as "safe guard", planets like Proxima Centauri b, also LHS~1440b and TRAPPIST-1g can have their equator-to-pole-wards stratosphere circulation being disrupted by equatorial wind jets in the stratosphere and Anti-Brewer-Dobson-circulation. The only available option to 'save' day side-wide stratosphere circulation on Proxima Centauri~b and similar planets is  deep stratosphere wind breaking. Future models of Proxima Centauri~b thus potentially need to take into account two scenarios when considering photo-chemistry and aerosol distribution in the stratosphere: an equatorial transport belt and day side-wide transport by thermally driven circulation.

Tidally locked ExoEarths with comparatively large orbital periods ($P_{orb} \geq 25$~days) like GJ~667 C e are completely safe from the tropical Rossby wave that cannot form efficient standing waves in this rotation regime. Thermally driven circulation is then efficiently transporting air masses in the stratosphere from the substellar point radially away towards the night side. This circulation can potentially lead to day side-wide distribution of ozone in the stratosphere, as indicated by \citet{Proedrou2016} but for a very slowly rotating ($P_{orb}=365$~days) tidally locked ExoEarth. Future work is needed to evaluate how the ozone content and distribution changes for more faster rotating tidally locked planets in the habitable zone of red dwarfs, e.g., with orbital periods between 30-40~days. Furthermore, the inherent time variability of ozone formation in planets around red dwarfs needs to be addressed in a 3D framework, extending work by \citet{Stelzer2013,Grenfell2014}. It will be also interesting to reevaluate in a 3D context the statement of \cite{Segura2010} that even a flaring star may not present a direct hazard to surface life.

In any case, it is not justified to assume Earth-like characteristics for tidally locked ExoEarths. Their stratosphere circulation  is very different from Earth, which does not show equatorial superrotation in the stratosphere and exhibits wave-driven (not thermally driven) stratosphere circulation - at least for now.

If stratosphere wind breaking is somehow diminished, we find in our model frame work that even Earth itself may develop equatorial wind jets in the stratosphere. This result is not completely outlandish: \citet{Pierrehumbert2000, Caballero2010} also argue that Earth may develop a stratospheric equatorial wind jet in the future due to global warming.  Something similar was found in \citet{Carone2016}  for tidally locked ExoEarths: Also there, higher temperatures lead to a strengthening of the tropical Rossby wave with either the appearance or strengthening of equatorial westward wind jets. Such winds can have potentially adverse effects on stratosphere circulation on Earth \citep{Korty2007}. The study of habitability on rocky exoplanets thus increases also our understanding of habitability on our own home world.

\section{Conclusion and outlook}
Our work generally emphasizes the importance of understanding the basic climate dynamics of habitable planets to guide a) the application of more complex, sophisticated climate models like those of \citet{Proedrou2016,Wolf2017,Kopparapu2017, Turbet2017, Boutle2017}, b) observations of biosignatures \citep{Barstow2016,Kreidberg2016} and c) evaluations of surface habitability \citep{Omalley2017,Segura2010}.  It is a first step to understand  how 3D transport may affect photo-chemistry of tidally locked exoplanets around ultra-cool dwarf stars for a wide range of relevant circulation scenarios - from very short ($P_{orb}=1.51$~days) to large ($P_{orb}=39$~days) orbital periods. Next, the production and transport of photo-chemically produced species needs to be investigated in more detail with a photo-chemical model coupled to our 3D circulation model to gain better insights about the prospects of detecting, e.g., ozone around very red dwarf stars like TRAPPIST-1.

Clearly, understanding the connection between photo-chemistry and stratosphere circulation on alien ExoEarths is one of the necessary steps to take before we can even start to infer habitability from observations. What is more, understanding stratosphere circulation on alien ExoEarths may even help to assess the limits of habitability on the one planet of which we definitely know that it is currently habitable: Earth.

\section*{Acknowledgments}
We acknowledge support from the KU Leuven projects IDO/10/2013 and GOA/2015-014 (2014-2018 KU Leuven). The computational resources and services used in this work were provided by the VSC (Flemish Supercomputer Center), funded by the Hercules Foundation and the Flemish Government - department EWI. LD acknowledges support from the ERC consolidator grant 646758 AEROSOL and the FWO Research Project grant G086217N.  LC thanks Patrick Barth for useful discussions. We also would like to thank the anonymous referee for her/his insightful comments that greatly helped to improve this manuscript.

\bibliographystyle{mnras}
\bibliography{Trappist} 

\begin{thebibliography}{}
\makeatletter
\relax
\def\mn@urlcharsother{\let\do\@makeother \do\$\do\&\do\#\do\^\do\_\do\%\do\~}
\def\mn@doi{\begingroup\mn@urlcharsother \@ifnextchar [ {\mn@doi@}
  {\mn@doi@[]}}
\def\mn@doi@[#1]#2{\def\@tempa{#1}\ifx\@tempa\@empty \href
  {http://dx.doi.org/#2} {doi:#2}\else \href {http://dx.doi.org/#2} {#1}\fi
  \endgroup}
\def\mn@eprint#1#2{\mn@eprint@#1:#2::\@nil}
\def\mn@eprint@arXiv#1{\href {http://arxiv.org/abs/#1} {{\tt arXiv:#1}}}
\def\mn@eprint@dblp#1{\href {http://dblp.uni-trier.de/rec/bibtex/#1.xml}
  {dblp:#1}}
\def\mn@eprint@#1:#2:#3:#4\@nil{\def\@tempa {#1}\def\@tempb {#2}\def\@tempc
  {#3}\ifx \@tempc \@empty \let \@tempc \@tempb \let \@tempb \@tempa \fi \ifx
  \@tempb \@empty \def\@tempb {arXiv}\fi \@ifundefined
  {mn@eprint@\@tempb}{\@tempb:\@tempc}{\expandafter \expandafter \csname
  mn@eprint@\@tempb\endcsname \expandafter{\@tempc}}}

\bibitem[\protect\citeauthoryear{{Abe}, {Abe-Ouchi}, {Sleep}  \&
  {Zahnle}}{{Abe} et~al.}{2011}]{Abe2011}
{Abe} Y.,  {Abe-Ouchi} A.,  {Sleep} N.~H.,   {Zahnle} K.~J.,  2011, \mn@doi
  [Astrobiology] {10.1089/ast.2010.0545}, \href
  {http://adsabs.harvard.edu/abs/2011AsBio..11..443A} {11, 443}

\bibitem[\protect\citeauthoryear{{Adcroft}, {Campin}, {Hill}  \&
  {Marshall}}{{Adcroft} et~al.}{2004}]{Adcroft2004}
{Adcroft} A.,  {Campin} J.-M.,  {Hill} C.,   {Marshall} J.,  2004, \mn@doi
  [Monthly Weather Review] {10.1175/MWR2823.1}, \href
  {http://adsabs.harvard.edu/abs/2004MWRv..132.2845A} {132, 2845}

\bibitem[\protect\citeauthoryear{{Anglada-Escud{\'e}}
  et~al.,}{{Anglada-Escud{\'e}} et~al.}{2013}]{Anglada2013}
{Anglada-Escud{\'e}} G.,  et~al., 2013, \mn@doi [\aap]
  {10.1051/0004-6361/201321331}, \href
  {http://adsabs.harvard.edu/abs/2013A%26A...556A.126A} {556, A126}

\bibitem[\protect\citeauthoryear{{Anglada-Escud{\'e}}
  et~al.,}{{Anglada-Escud{\'e}} et~al.}{2016}]{Anglada2016}
{Anglada-Escud{\'e}} G.,  et~al., 2016, \mn@doi [\nat] {10.1038/nature19106},
  \href {http://adsabs.harvard.edu/abs/2016Natur.536..437A} {536, 437}

\bibitem[\protect\citeauthoryear{{Barstow} \& {Irwin}}{{Barstow} \&
  {Irwin}}{2016}]{Barstow2016}
{Barstow} J.~K.,  {Irwin} P.~G.~J.,  2016, \mn@doi [\mnras]
  {10.1093/mnrasl/slw109}, \href
  {http://adsabs.harvard.edu/abs/2016MNRAS.461L..92B} {461, L92}

\bibitem[\protect\citeauthoryear{{Bolmont}, {Selsis}, {Owen}, {Ribas},
  {Raymond}, {Leconte}  \& {Gillon}}{{Bolmont} et~al.}{2017}]{Bolmont2017}
{Bolmont} E.,  {Selsis} F.,  {Owen} J.~E.,  {Ribas} I.,  {Raymond} S.~N.,
  {Leconte} J.,   {Gillon} M.,  2017, \mn@doi [\mnras] {10.1093/mnras/stw2578},
  \href {http://adsabs.harvard.edu/abs/2017MNRAS.464.3728B} {464, 3728}

\bibitem[\protect\citeauthoryear{{Boutle}, {Mayne}, {Drummond}, {Manners},
  {Goyal}, {Hugo Lambert}, {Acreman}  \& {Earnshaw}}{{Boutle}
  et~al.}{2017}]{Boutle2017}
{Boutle} I.~A.,  {Mayne} N.~J.,  {Drummond} B.,  {Manners} J.,  {Goyal} J.,
  {Hugo Lambert} F.,  {Acreman} D.~M.,   {Earnshaw} P.~D.,  2017, \mn@doi
  [\aap] {10.1051/0004-6361/201630020}, \href
  {http://adsabs.harvard.edu/abs/2017A%26A...601A.120B} {601, A120}

\bibitem[\protect\citeauthoryear{{Brewer}}{{Brewer}}{1949}]{Brewer1949}
{Brewer} A.~W.,  1949, \mn@doi [Quarterly Journal of the Royal Meteorological
  Society] {10.1002/qj.49707532603}, \href
  {http://adsabs.harvard.edu/abs/1949QJRMS..75..351B} {75, 351}

\bibitem[\protect\citeauthoryear{{Caballero} \& {Huber}}{{Caballero} \&
  {Huber}}{2010}]{Caballero2010}
{Caballero} R.,  {Huber} M.,  2010, \mn@doi [\grl] {10.1029/2010GL043468},
  \href {http://adsabs.harvard.edu/abs/2010GeoRL..3711701C} {37, L11701}

\bibitem[\protect\citeauthoryear{{Carone}, {Keppens}  \& {Decin}}{{Carone}
  et~al.}{2014}]{Carone2014}
{Carone} L.,  {Keppens} R.,   {Decin} L.,  2014, \mn@doi [\mnras]
  {10.1093/mnras/stu1793}, \href
  {http://adsabs.harvard.edu/abs/2014MNRAS.445..930C} {445, 930}

\bibitem[\protect\citeauthoryear{{Carone}, {Keppens}  \& {Decin}}{{Carone}
  et~al.}{2015}]{Carone2015}
{Carone} L.,  {Keppens} R.,   {Decin} L.,  2015, \mn@doi [\mnras]
  {10.1093/mnras/stv1752}, \href
  {http://adsabs.harvard.edu/abs/2015MNRAS.453.2412C} {453, 2412}

\bibitem[\protect\citeauthoryear{{Carone}, {Keppens}  \& {Decin}}{{Carone}
  et~al.}{2016}]{Carone2016}
{Carone} L.,  {Keppens} R.,   {Decin} L.,  2016, \mn@doi [\mnras]
  {10.1093/mnras/stw1265}, \href
  {http://adsabs.harvard.edu/abs/2016MNRAS.461.1981C} {461, 1981}

\bibitem[\protect\citeauthoryear{{Cordero} \& {Kawa}}{{Cordero} \&
  {Kawa}}{2001}]{Cordero2001}
{Cordero} E.~C.,  {Kawa} S.~R.,  2001, \mn@doi [\jgr] {10.1029/2001JD900004},
  \href {http://adsabs.harvard.edu/abs/2001JGR...10612227C} {106, 12}

\bibitem[\protect\citeauthoryear{{Deleuil} et~al.,}{{Deleuil}
  et~al.}{2008}]{Deleuil2008}
{Deleuil} M.,  et~al., 2008, \mn@doi [\aap] {10.1051/0004-6361:200810625},
  \href {http://adsabs.harvard.edu/abs/2008A%26A...491..889D} {491, 889}

\bibitem[\protect\citeauthoryear{{Delfosse} et~al.,}{{Delfosse}
  et~al.}{2013}]{Delfosse2013}
{Delfosse} X.,  et~al., 2013, \mn@doi [\aap] {10.1051/0004-6361/201219013},
  \href {http://adsabs.harvard.edu/abs/2013A%26A...553A...8D} {553, A8}

\bibitem[\protect\citeauthoryear{{Dittmann} et~al.,}{{Dittmann}
  et~al.}{2017}]{Dittmann2017}
{Dittmann} J.~A.,  et~al., 2017, \mn@doi [\nat] {10.1038/nature22055}, \href
  {http://adsabs.harvard.edu/abs/2017Natur.544..333D} {544, 333}

\bibitem[\protect\citeauthoryear{{Dobson}}{{Dobson}}{1931}]{Dobson1931}
{Dobson} G.~M.~B.,  1931, \mn@doi [\nat] {10.1038/127668a0}, \href
  {http://adsabs.harvard.edu/abs/1931Natur.127..668D} {127, 668}

\bibitem[\protect\citeauthoryear{{Edson}, {Lee}, {Bannon}, {Kasting}  \&
  {Pollard}}{{Edson} et~al.}{2011}]{Edson2011}
{Edson} A.,  {Lee} S.,  {Bannon} P.,  {Kasting} J.~F.,   {Pollard} D.,  2011,
  \mn@doi [\icarus] {10.1016/j.icarus.2010.11.023}, \href
  {http://adsabs.harvard.edu/abs/2011Icar..212....1E} {212, 1}

\bibitem[\protect\citeauthoryear{{Fulton} et~al.,}{{Fulton}
  et~al.}{2017}]{Fulton2017}
{Fulton} B.~J.,  et~al., 2017, preprint, \href
  {http://adsabs.harvard.edu/abs/2017arXiv170310375F} {} (\mn@eprint {arXiv}
  {1703.10375})

\bibitem[\protect\citeauthoryear{{Gillon} et~al.,}{{Gillon}
  et~al.}{2016}]{Gillon2016}
{Gillon} M.,  et~al., 2016, \mn@doi [\nat] {10.1038/nature17448}, \href
  {http://adsabs.harvard.edu/abs/2016Natur.533..221G} {533, 221}

\bibitem[\protect\citeauthoryear{{Gillon} et~al.,}{{Gillon}
  et~al.}{2017}]{Gillon2017}
{Gillon} M.,  et~al., 2017, \mn@doi [\nat] {10.1038/nature21360}, \href
  {http://adsabs.harvard.edu/abs/2017Natur.542..456G} {542, 456}

\bibitem[\protect\citeauthoryear{{Grenfell}, {Gebauer}, {v.~Paris}, {Godolt}
  \& {Rauer}}{{Grenfell} et~al.}{2014}]{Grenfell2014}
{Grenfell} J.~L.,  {Gebauer} S.,  {v.~Paris} P.,  {Godolt} M.,   {Rauer} H.,
  2014, \mn@doi [\planss] {10.1016/j.pss.2013.10.006}, \href
  {http://adsabs.harvard.edu/abs/2014P%26SS...98...66G} {98, 66}

\bibitem[\protect\citeauthoryear{{Haynes}}{{Haynes}}{2005}]{Haynes2005}
{Haynes} P.,  2005, \mn@doi [Annual Review of Fluid Mechanics]
  {10.1146/annurev.fluid.37.061903.175710}, \href
  {http://adsabs.harvard.edu/abs/2005AnRFM..37..263H} {37, 263}

\bibitem[\protect\citeauthoryear{{Held} \& {Hou}}{{Held} \&
  {Hou}}{1980}]{Held1980}
{Held} I.~M.,  {Hou} A.~Y.,  1980, \mn@doi [Journal of Atmospheric Sciences]
  {10.1175/1520-0469(1980)037<0515:NASCIA>2.0.CO;2}, \href
  {http://adsabs.harvard.edu/abs/1980JAtS...37..515H} {37, 515}

\bibitem[\protect\citeauthoryear{{Held} \& {Suarez}}{{Held} \&
  {Suarez}}{1994}]{Held1994}
{Held} I.~M.,  {Suarez} M.~J.,  1994, \mn@doi [Bulletin of the American
  Meteorological Society] {10.1175/1520-0477(1994)075<1825:APFTIO>2.0.CO;2},
  \href {http://adsabs.harvard.edu/abs/1994BAMS...75.1825H} {75, 1825}

\bibitem[\protect\citeauthoryear{{Holton}}{{Holton}}{1992}]{Holton}
{Holton} J.~R.,  1992, {An introduction to dynamic meteorology}, 3 edn.
San Diego, New York: Academic Press

\bibitem[\protect\citeauthoryear{{Kopparapu} et~al.,}{{Kopparapu}
  et~al.}{2013}]{Kopparapu2013}
{Kopparapu} R.~K.,  et~al., 2013, \mn@doi [\apj] {10.1088/0004-637X/765/2/131},
  \href {http://adsabs.harvard.edu/abs/2013ApJ...765..131K} {765, 131}

\bibitem[\protect\citeauthoryear{{Kopparapu}, {Wolf}, {Arney}, {Batalha},
  {Haqq-Misra}, {Grimm}  \& {Heng}}{{Kopparapu} et~al.}{2017}]{Kopparapu2017}
{Kopparapu} R.~k.,  {Wolf} E.~T.,  {Arney} G.,  {Batalha} N.~E.,  {Haqq-Misra}
  J.,  {Grimm} S.~L.,   {Heng} K.,  2017, \mn@doi [\apj]
  {10.3847/1538-4357/aa7cf9}, \href
  {http://adsabs.harvard.edu/abs/2017ApJ...845....5K} {845, 5}

\bibitem[\protect\citeauthoryear{{Korty} \& {Emanuel}}{{Korty} \&
  {Emanuel}}{2007}]{Korty2007}
{Korty} R.~L.,  {Emanuel} K.~A.,  2007, \mn@doi [Journal of Climate]
  {10.1175/2007JCLI1556.1}, \href
  {http://adsabs.harvard.edu/abs/2007JCli...20.5213K} {20, 5213}

\bibitem[\protect\citeauthoryear{{Kreidberg} \& {Loeb}}{{Kreidberg} \&
  {Loeb}}{2016}]{Kreidberg2016}
{Kreidberg} L.,  {Loeb} A.,  2016, \mn@doi [\apjl]
  {10.3847/2041-8205/832/1/L12}, \href
  {http://adsabs.harvard.edu/abs/2016ApJ...832L..12K} {832, L12}

\bibitem[\protect\citeauthoryear{{Luger} et~al.,}{{Luger}
  et~al.}{2017}]{Luger2017}
{Luger} R.,  et~al., 2017, \mn@doi [Nature Astronomy]
  {10.1038/s41550-017-0129}, \href
  {http://adsabs.harvard.edu/abs/2017NatAs...1E.129L} {1, 0129}

\bibitem[\protect\citeauthoryear{{Manabe} \& {Hunt}}{{Manabe} \&
  {Hunt}}{1968}]{Manabe1968}
{Manabe} S.,  {Hunt} B.~G.,  1968, \mn@doi [Monthly Weather Review]
  {10.1175/1520-0493(1968)096<0477:EWASGC>2.0.CO;2}, \href
  {http://adsabs.harvard.edu/abs/1968MWRv...96..477M} {96, 477}

\bibitem[\protect\citeauthoryear{{Montmessin} \& {Lef{\`e}vre}}{{Montmessin} \&
  {Lef{\`e}vre}}{2013}]{Mont2013}
{Montmessin} F.,  {Lef{\`e}vre} F.,  2013, \mn@doi [Nature Geoscience]
  {10.1038/ngeo1957}, \href {http://adsabs.harvard.edu/abs/2013NatGe...6..930M}
  {6, 930}

\bibitem[\protect\citeauthoryear{{Morley}, {Kreidberg}, {Rustamkulov},
  {Robinson}  \& {Fortney}}{{Morley} et~al.}{2017}]{Morley2017}
{Morley} C.~V.,  {Kreidberg} L.,  {Rustamkulov} Z.,  {Robinson} T.,   {Fortney}
  J.~J.,  2017, preprint, \href
  {http://adsabs.harvard.edu/abs/2017arXiv170804239M} {} (\mn@eprint {arXiv}
  {1708.04239})

\bibitem[\protect\citeauthoryear{{O'Malley-James} \&
  {Kaltenegger}}{{O'Malley-James} \& {Kaltenegger}}{2017}]{Omalley2017}
{O'Malley-James} J.~T.,  {Kaltenegger} L.,  2017, \mn@doi [\mnras]
  {10.1093/mnrasl/slx047}, \href
  {http://adsabs.harvard.edu/abs/2017MNRAS.469L..26O} {469, L26}

\bibitem[\protect\citeauthoryear{{Pierrehumbert}}{{Pierrehumbert}}{2000}]{Pierrehumbert2000}
{Pierrehumbert} R.~T.,  2000, Proceedings of the National Academy of Science,
  \href {http://adsabs.harvard.edu/abs/2000PNAS...97.1355P} {97, 1355}

\bibitem[\protect\citeauthoryear{{Prinn} \& {Fegley}}{{Prinn} \&
  {Fegley}}{1987}]{Prinn1987}
{Prinn} R.~G.,  {Fegley} B.,  1987, \mn@doi [Annual Review of Earth and
  Planetary Sciences] {10.1146/annurev.ea.15.050187.001131}, \href
  {http://adsabs.harvard.edu/abs/1987AREPS..15..171P} {15, 171}

\bibitem[\protect\citeauthoryear{{Proedrou} \& {Hocke}}{{Proedrou} \&
  {Hocke}}{2016}]{Proedrou2016}
{Proedrou} E.,  {Hocke} K.,  2016, \mn@doi [Earth, Planets, and Space]
  {10.1186/s40623-016-0461-x}, \href
  {http://adsabs.harvard.edu/abs/2016EP%26S...68...96P} {68, 96}

\bibitem[\protect\citeauthoryear{{Quirrenbach}, {Amado}, {Mandel}, {Caballero},
  {Ribas}, {Reiners}, {Mundt}  \& {CARMENES Consortium}}{{Quirrenbach}
  et~al.}{2010}]{Quirrenbach2010}
{Quirrenbach} A.,  {Amado} P.~J.,  {Mandel} H.,  {Caballero} J.~A.,  {Ribas}
  I.,  {Reiners} A.,  {Mundt} R.,   {CARMENES Consortium} 2010, in {Coud{\'e}
  du Foresto} V.,  {Gelino} D.~M.,   {Ribas} I.,  eds,  Astronomical Society of
  the Pacific Conference Series Vol. 430, Pathways Towards Habitable Planets.
  p.~521 (\mn@eprint {arXiv} {0912.0561})

\bibitem[\protect\citeauthoryear{{Segura}, {Kasting}, {Meadows}, {Cohen},
  {Scalo}, {Crisp}, {Butler}  \& {Tinetti}}{{Segura} et~al.}{2005}]{Segura2005}
{Segura} A.,  {Kasting} J.~F.,  {Meadows} V.,  {Cohen} M.,  {Scalo} J.,
  {Crisp} D.,  {Butler} R.~A.~H.,   {Tinetti} G.,  2005, \mn@doi [Astrobiology]
  {10.1089/ast.2005.5.706}, \href
  {http://adsabs.harvard.edu/abs/2005AsBio...5..706S} {5, 706}

\bibitem[\protect\citeauthoryear{{Segura}, {Walkowicz}, {Meadows}, {Kasting}
  \& {Hawley}}{{Segura} et~al.}{2010}]{Segura2010}
{Segura} A.,  {Walkowicz} L.~M.,  {Meadows} V.,  {Kasting} J.,   {Hawley} S.,
  2010, \mn@doi [Astrobiology] {10.1089/ast.2009.0376}, \href
  {http://adsabs.harvard.edu/abs/2010AsBio..10..751S} {10, 751}

\bibitem[\protect\citeauthoryear{{Shaw} \& {Shepherd}}{{Shaw} \&
  {Shepherd}}{2008}]{Shaw2008}
{Shaw} T.~A.,  {Shepherd} T.~G.,  2008, \mn@doi [Nature Geoscience]
  {10.1038/ngeo.2007.53}, \href
  {http://adsabs.harvard.edu/abs/2008NatGe...1...12S} {1, 12}

\bibitem[\protect\citeauthoryear{{Showman} \& {Kaspi}}{{Showman} \&
  {Kaspi}}{2013}]{Showman2013}
{Showman} A.~P.,  {Kaspi} Y.,  2013, \mn@doi [\apj]
  {10.1088/0004-637X/776/2/85}, \href
  {http://adsabs.harvard.edu/abs/2013ApJ...776...85S} {776, 85}

\bibitem[\protect\citeauthoryear{{Showman} \& {Polvani}}{{Showman} \&
  {Polvani}}{2011}]{Showman2011}
{Showman} A.~P.,  {Polvani} L.~M.,  2011, \mn@doi [\apj]
  {10.1088/0004-637X/738/1/71}, \href
  {http://adsabs.harvard.edu/abs/2011ApJ...738...71S} {738, 71}

\bibitem[\protect\citeauthoryear{{Showman}, {Wordsworth}, {Merlis}  \&
  {Kaspi}}{{Showman} et~al.}{2013}]{Showmanbook}
{Showman} A.~P.,  {Wordsworth} R.~D.,  {Merlis} T.~M.,   {Kaspi} Y.,  2013,
  {Atmospheric Circulation of Terrestrial Exoplanets}.
pp 277--326, \mn@doi{10.2458/azu_uapress_9780816530595-ch12}

\bibitem[\protect\citeauthoryear{{Stelzer}, {Marino}, {Micela},
  {L{\'o}pez-Santiago}  \& {Liefke}}{{Stelzer} et~al.}{2013}]{Stelzer2013}
{Stelzer} B.,  {Marino} A.,  {Micela} G.,  {L{\'o}pez-Santiago} J.,   {Liefke}
  C.,  2013, \mn@doi [\mnras] {10.1093/mnras/stt225}, \href
  {http://adsabs.harvard.edu/abs/2013MNRAS.431.2063S} {431, 2063}

\bibitem[\protect\citeauthoryear{{Turbet} et~al.,}{{Turbet}
  et~al.}{2017}]{Turbet2017}
{Turbet} M.,  et~al., 2017, preprint, \href
  {http://adsabs.harvard.edu/abs/2017arXiv170706927T} {} (\mn@eprint {arXiv}
  {1707.06927})

\bibitem[\protect\citeauthoryear{{Vallis}}{{Vallis}}{2006}]{Vallis}
{Vallis} G.~K.,  2006, {Atmospheric and Oceanic Fluid Dynamics}.
Cambridge University Press, \mn@doi{10.2277/0521849691}

\bibitem[\protect\citeauthoryear{{Venot}, {Rocchetto}, {Carl}, {Roshni Hashim}
  \& {Decin}}{{Venot} et~al.}{2016}]{Venot2016}
{Venot} O.,  {Rocchetto} M.,  {Carl} S.,  {Roshni Hashim} A.,   {Decin} L.,
  2016, \mn@doi [\apj] {10.3847/0004-637X/830/2/77}, \href
  {http://adsabs.harvard.edu/abs/2016ApJ...830...77V} {830, 77}

\bibitem[\protect\citeauthoryear{{Vida}, {K{\H o}v{\'a}ri}, {P{\'a}l},
  {Ol{\'a}h}  \& {Kriskovics}}{{Vida} et~al.}{2017}]{Vida2017}
{Vida} K.,  {K{\H o}v{\'a}ri} Z.,  {P{\'a}l} A.,  {Ol{\'a}h} K.,   {Kriskovics}
  L.,  2017, \mn@doi [\apj] {10.3847/1538-4357/aa6f05}, \href
  {http://adsabs.harvard.edu/abs/2017ApJ...841..124V} {841, 124}

\bibitem[\protect\citeauthoryear{{Wang}, {Wu}, {Barclay}  \& {Laughlin}}{{Wang}
  et~al.}{2017}]{Wang2017}
{Wang} S.,  {Wu} D.-H.,  {Barclay} T.,   {Laughlin} G.~P.,  2017, preprint,
  \href {http://adsabs.harvard.edu/abs/2017arXiv170404290W} {} (\mn@eprint
  {arXiv} {1704.04290})

\bibitem[\protect\citeauthoryear{{Williamson}, {Olson}  \&
  {Boville}}{{Williamson} et~al.}{1998}]{William1998}
{Williamson} D.~L.,  {Olson} J.~G.,   {Boville} B.~A.,  1998, \mn@doi [Monthly
  Weather Review] {10.1175/1520-0493(1998)126<1001:ACOSLA>2.0.CO;2}, \href
  {http://adsabs.harvard.edu/abs/1998MWRv..126.1001W} {126, 1001}

\bibitem[\protect\citeauthoryear{{Wolf}}{{Wolf}}{2017}]{Wolf2017}
{Wolf} E.~T.,  2017, \mn@doi [\apjl] {10.3847/2041-8213/aa693a}, \href
  {http://adsabs.harvard.edu/abs/2017ApJ...839L...1W} {839, L1}

\bibitem[\protect\citeauthoryear{{Yang}, {Cowan}  \& {Abbot}}{{Yang}
  et~al.}{2013}]{Yang2013}
{Yang} J.,  {Cowan} N.~B.,   {Abbot} D.~S.,  2013, \mn@doi [\apjl]
  {10.1088/2041-8205/771/2/L45}, \href
  {http://adsabs.harvard.edu/abs/2013ApJ...771L..45Y} {771, L45}

\bibitem[\protect\citeauthoryear{{Yang}, {Bou{\'e}}, {Fabrycky}  \&
  {Abbot}}{{Yang} et~al.}{2014}]{Yang2014}
{Yang} J.,  {Bou{\'e}} G.,  {Fabrycky} D.~C.,   {Abbot} D.~S.,  2014, \mn@doi
  [\apjl] {10.1088/2041-8205/787/1/L2}, \href
  {http://adsabs.harvard.edu/abs/2014ApJ...787L...2Y} {787, L2}

\bibitem[\protect\citeauthoryear{{Zsom}, {Seager}, {de Wit}  \&
  {Stamenkovi{\'c}}}{{Zsom} et~al.}{2013}]{Zsom2013}
{Zsom} A.,  {Seager} S.,  {de Wit} J.,   {Stamenkovi{\'c}} V.,  2013, \mn@doi
  [\apj] {10.1088/0004-637X/778/2/109}, \href
  {http://adsabs.harvard.edu/abs/2013ApJ...778..109Z} {778, 109}

\bibitem[\protect\citeauthoryear{{Zuluaga} \& {Bustamante}}{{Zuluaga} \&
  {Bustamante}}{2016}]{Zuluaga2016}
{Zuluaga} J.~I.,  {Bustamante} S.,  2016, preprint, \href
  {http://adsabs.harvard.edu/abs/2016arXiv160900707Z} {} (\mn@eprint {arXiv}
  {1609.00707})

\makeatother
\end{thebibliography}
\label{lastpage}
\end{document}